\documentclass[sigconf]{acmart}
 \usepackage{graphicx}
\usepackage{color}	
\usepackage{subcaption}
\usepackage{caption}
 \usepackage[utf8]{inputenc} 
 \usepackage{amsmath}
 \usepackage{algorithm}
\usepackage{algorithmic}
 
\begin{document}
%10 pages.

\title{Peak Transmission Rate Resilient Crosslayer Broadcast for Body Area Networks} %Reliable Native  Broadcast 
\thanks{This work was funded by SMART-BAN project (Labex SMART) http://www.smart-labex.fr.}

\acmConference[MobiHoc'17]{ACM Conference}{July 2017}{Madras, Chennai, India}

\author{Wafa Badreddine}
\affiliation{UPMC Sorbonne Universites, \\
LIP6-CNRS UMR 7606, France}
\email{wafa.badreddine@lip6.fr}

\author{Maria Potop-Butucaru}
\affiliation{UPMC Sorbonne Universites, \\
LIP6-CNRS UMR 7606, France}
\email{maria.potop-butucaru@lip6.fr}

\begin{abstract}
Wireless Body Area Networks (WBAN) open an interdisciplinary area within Wireless Sensor Networks (WSN) research, with a tremendous impact in healthcare area where sensors are used to monitor, collect and transmit biological parameters of the human body. We propose the first network-MAC cross-layer broadcast protocol in WBAN. Our protocol, evaluated in the OMNET++ simulator enriched with realistic human body mobility models and channel models issued from the recent research on biomedical and health informatics, outperforms existing flat broadcast strategies in terms of percentage of covered nodes, energy consumption and correct reception of causally-ordered packets (i.e. packets are received in the same order as they were sent by the sink). Furthermore, we investigate the resilience of both existing flat broadcast strategies and our new protocol face to various transmission rates and human body mobility. Existing flat broadcast strategies without exception start to have a drastic drop of performances for transmission rates above $11$Kb/s while our cross-layer protocol performances maintains its good performances for transmission rates up to $190$Kb/s.

%$20$ packets per seconds while our cross-layer protocol maintains its good performances up to $350$ packets per second.

%In this paper, we present \emph{CLPB} for \textbf{Cross Layer Protocol for Broadcast} with the particularity of handling broadcast in WBAN. 
%\emph{CLPB} features are to handle \emph{Sink} traffic requirement and to adapt to human body mobility. Thus \emph{CLPB} is a slotted protocol that involves tree construction to both control medium access and broadcast process. \emph{CLPB} strengths are light synchronization to ensure communication, no control packets exchange and unnecessary data packets retransmissions. Our protocol evaluation is twofold. First, \emph{CLPB} is compared to other broadcast strategies via simulations on Omnet++ simulator and Mixim framework. Second, we studied \emph{CLPB} causal order resiliency. Thus, we stressed our protocol with transmission rate up to $1000$ packets per second. Results show that \emph{CLPB} performs well up to $350$ packets per second and outperforms old broadcast strategies based on IEEE 802.15.4 that ensure good performance up to $20$ packets per second in average.
%%Our new protocol consumes $9$ time less energy than the other broadcast strategies.
 
\end{abstract} 

\maketitle

\section{Introduction} 
In healthcare area WBAN (Wireless Body Area Networks) (\cite{WBAN1}, \cite{Latre:2011:SWB}, \cite{DBLP:journals/corr/abs-1303-2062}) emerged as a viable solution in response to the various disadvantages associated with wired sensors commonly used to monitor patients in hospitals and emergency rooms. WBAN is a promising technology and shall be increasingly necessary for monitoring, diagnosing and treating populations. Recent medical reports predict that the number of people using home health technologies will enormously increase from 14.3 to 78 million consumers from 2014 to 2020~\cite{32}, respectively. Additionally, body sensors shipments will hit 3.1 million units every year.

In WBAN tiny devices with low computing power and limited life, deployed in/on or around a human body, are able to detect and collect the physiological phenomena of the human body (such as: EEG, ECG, SpO2, lactic acid, etc.), and further transmit this information to a collector point (i.e \emph{Sink}) that will process it, take decisions, alert or record. 
 
WBANs \cite{6755575} differ from typical large-scale wireless sensor networks in many aspects: the size of the network is limited to a dozen of nodes, in-network mobility follows the body movements and the wireless channel has its specificities. Links have, in general, a very short range and a quality that varies with the wearer's posture, but remains low in the general case. Indeed, the transmission power is kept low, which improves devices autonomy and reduces wearers electromagnetic exposition. Consequently, the effects of body absorption, reflections and interference cannot be neglected and it is difficult to maintain a direct link (one-hop) between a data collection point and all WBAN nodes. Although, recent research \cite{channel} advocates for using \emph{multi-hop} communication in WBAN, very few multi-hop communication protocols have been proposed so far and even fewer are optimized for the human body mobility. 
% Although, recent research\cite{5}% 
 In this paper we are focusing \emph{multi-hop broadcast} protocols where the packets sent by a \emph{Sink} node have be received by all the nodes in the network. Additionally, we investigate the native capability of these protocols to handle the reception of packets in the order they have been sent by the \emph{Sink} without using heavy synchronization mechanisms.
 
\paragraph{Contributions} The current work extends in several ways the results in \cite{BCPP15} where the authors propose a collection of network layer broadcast strategies designed specifically to be resilient to human body mobility in WBANs. First, we stressed all the strategies proposed in \cite{BCPP15} with various transmission rates up to $544$Kb/s in seven different mobility postures (walk, run, sleep, weak walk, etc). Our evaluation has been conducted in the OMNET++ simulator that we enriched with realistic human body mobility models and channel models issued from the recent research on biomedical and health informatics \cite{channel}.
With no exception, the existing flat broadcast strategies register a dramatic drop of performances when the transmission rate is superior to $11$Kb/s. Second, we propose the first network-MAC layer broadcast protocol, \emph{CLBP}, designed for multi-hop topologies and resilient to realistic human body postures and mobility. Our protocol is optimized to exploit the human body mobility by carefully choosing the most reliable communication paths in each studied posture. Moreover, our protocol includes a slot assignment mechanism that reduces the energy consumption, collisions, idle listening and overhearing. Additionally, \emph{CLBP} includes a synchronization scheme that helps nodes to resynchronize with the sink on the fly.
 Our protocol outperforms existing flat broadcast strategies in terms of percentage of covered nodes, energy consumption and correct reception of causally-ordered packets (i.e. packets are received in the order of their sending). Furthermore, our protocol maintains its good performances up to $190$Kb/s transmission rates. 

\paragraph{Roadmap} 
This paper is organized as follows: Section \ref{RelatedWork} presents the related work. Section \ref{ChannelModel} presents the channel model and realistic human body mobility. In Section \ref{NewCrossProtocol} we detail CLPB our new cross layer broadcast protocol. In section \ref{sec:simulation}, we extensively evaluate protocols in \cite{BCPP15} and our new cross layer protocol. Section \ref{Conclusion} concludes the paper.
 
\section{Related Work} 
\label{RelatedWork}

Inspired by the tremendous work in WSN, adhoc networks\cite{Ni:1999sd} and DTN\cite{DTN1}, \cite{BCPP15} proposes for the first time in the WBAN
a set of multi-hop broadcast strategies and extensively evaluate them against realistic human body mobility. The evaluation focus the scenario where a specific node in the network, \emph{Sink} sends a single packet to all the nodes in the network. Our current work extends the study in \cite{BCPP15} by stressing the strategies proposed in \cite{BCPP15} with various transmission rates. Our results, show a drastic drop of performances when the transmission rates. In order to encompass this drawback we propose a new cross layer broadcast protocol that exploits the communication graphs defined by the body mobility in order to optimize the communication. Hence our new protocol outperform existing flat strategies in terms of end-to-end delay, percentage of covered nodes, energy and native capability to provide causal order reception of packets. 
%%%%%%%%%%%%%%%%%%%%%%%%%% Cross Layer
%Many works have exploited cross layer principle in different networks ranging from sensor to DTN. 
%Cross layer approach showed a good compromise between reliability, energy efficiency,etc.

%Cross Layer optimizations have been widely studied. For examples: channel-aware adhoc routing protocols, content-aware encoding of video streams, application-aware power management protocols.

There are very few cross-layered protocols specifically designed for WBAN (please refer to \cite{s121114730}, \cite{Ullah}, \cite{Ullah2012} and \cite{chica} for a review of cross layer concepts, methods and existing protocols overview). 

In the following we will discuss mainly multi-hop MAC-network cross-layer protocols that involve MAC and Network layers. These protocols focus \emph{converge-cast} (nodes send packets to the \emph{Sink}). To the best of our knowledge no cross-layer broadcast protocol has been proposed so far for multi-hop WBAN.

In \cite{4205278}, authors proposed WASP (implemented in \cite{WASPimplem}), a converge-cast cross-layer protocol. WASP is a slotted protocol that uses a spanning tree for medium access coordination and traffic routing. 
 \cite{4205278} is not resilient to realistic human body mobility. 
 
CICADA, \cite{CICADA}, was proposed as an improvement of WASP. CICADA aims at reducing energy consumption with the use of a spanning tree and an assignment of transmission slots. In addition, CICADA handles network nodes joining. CICADA has never been evaluated against realistic human body mobility and various transmission rates. Moreover, the medium access control scheme proposed in CICADA is specifically designed to handle converge-cast. Its adaptation to broadcast is an open question \cite{CICADA}.
 
\section{Channel model}
\label{ChannelModel}

We implemented the realistic channel model in \cite{channel} over the physical layer implementation provided by the Mixim framework \cite{mixim}. This channel model of an on-body $2.45\,GHz$ channel between $7$ nodes, that belong to the same WBAN, using small directional antennas modeled as if they were $1.5cm$ away from the body. Nodes are assumed to be attached to the human body on the head (2), chest (1), upper arm (3), wrist (6), navel (0), thigh (5), and ankle (4). 

\begin{figure}[htbp]
\centering
 \begin{subfigure}{0.9\columnwidth}
 \centering
 \includegraphics[width=\textwidth,height=2.7cm]{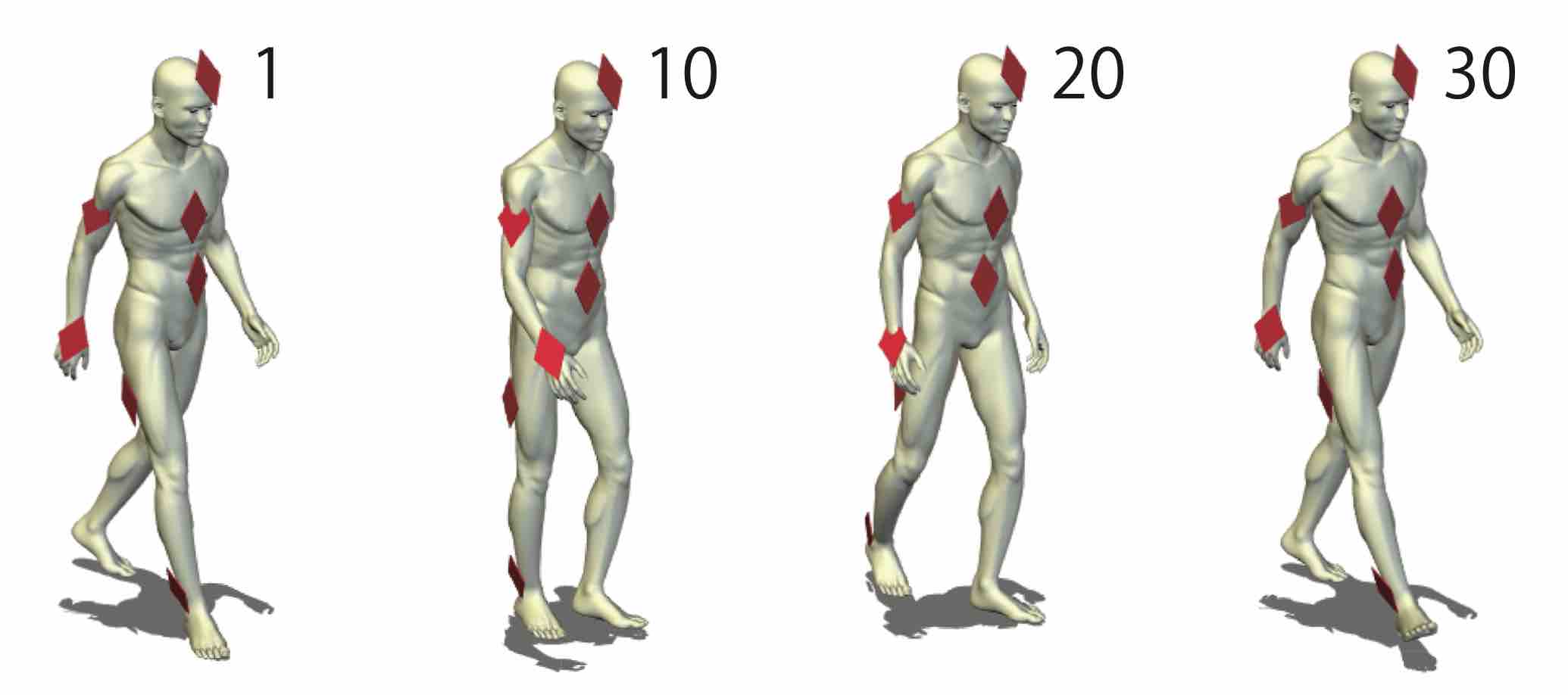}
 \caption{Walking}
 \end{subfigure}
 \begin{subfigure}{0.9\columnwidth}
 \centering
 \includegraphics[width=\textwidth,height=2.5cm]{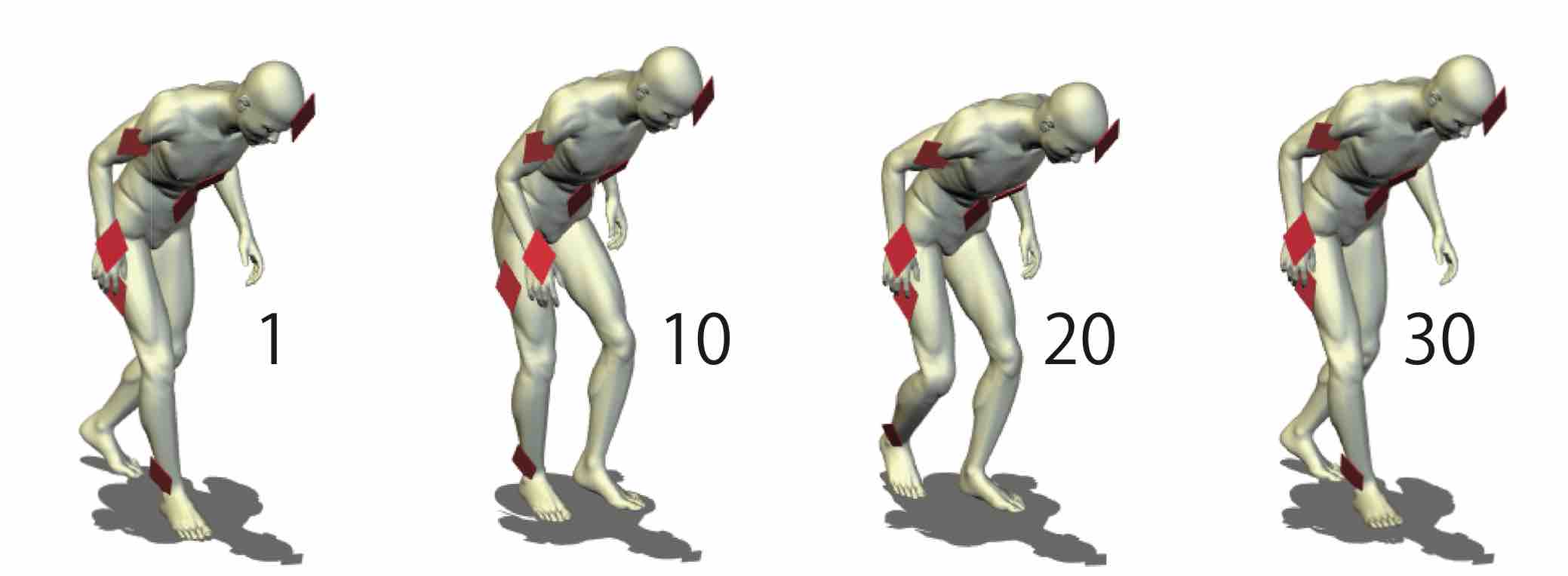}
 \caption{Walking weakly}
 \end{subfigure}
 \begin{subfigure}{0.9\columnwidth}
 \centering
 \includegraphics[width=\textwidth,height=2.5cm]{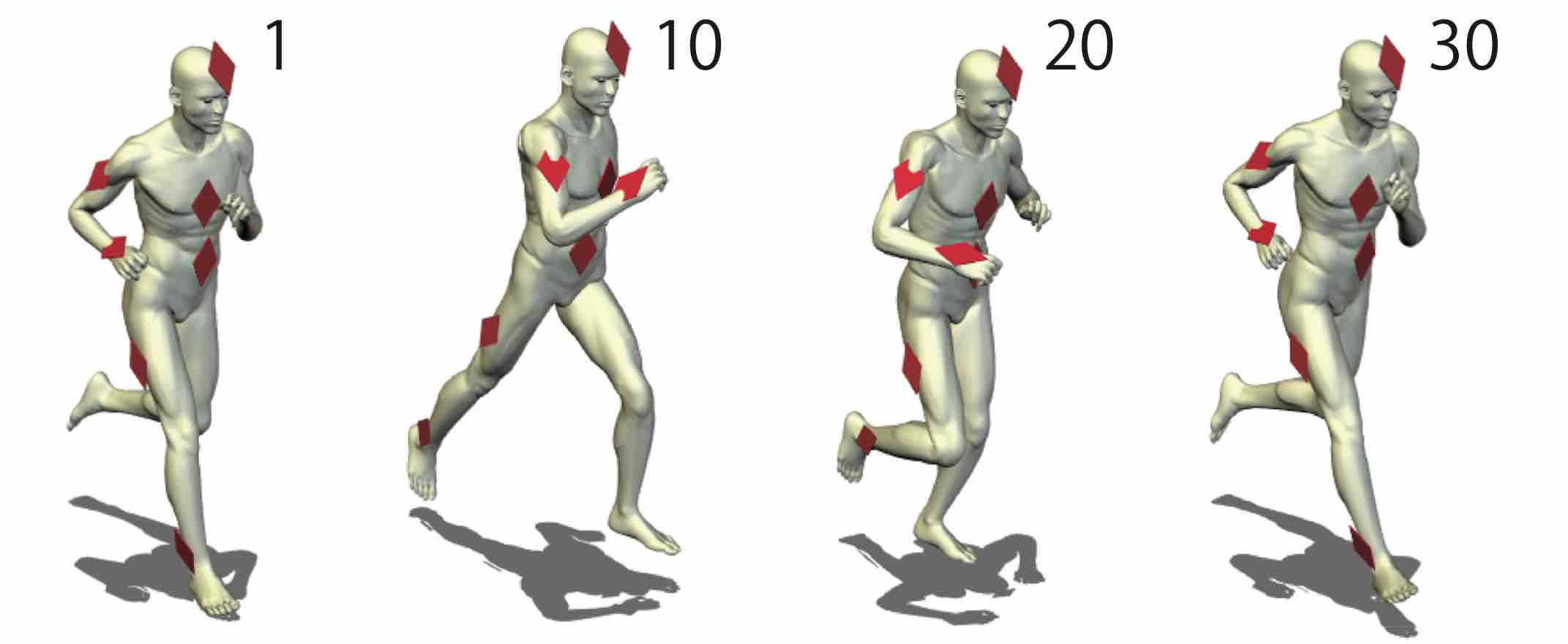}
 \caption{Running}
 \end{subfigure}
 \begin{subfigure}{0.9\columnwidth}
 \centering
 \includegraphics[width=\textwidth,height=2.5cm]{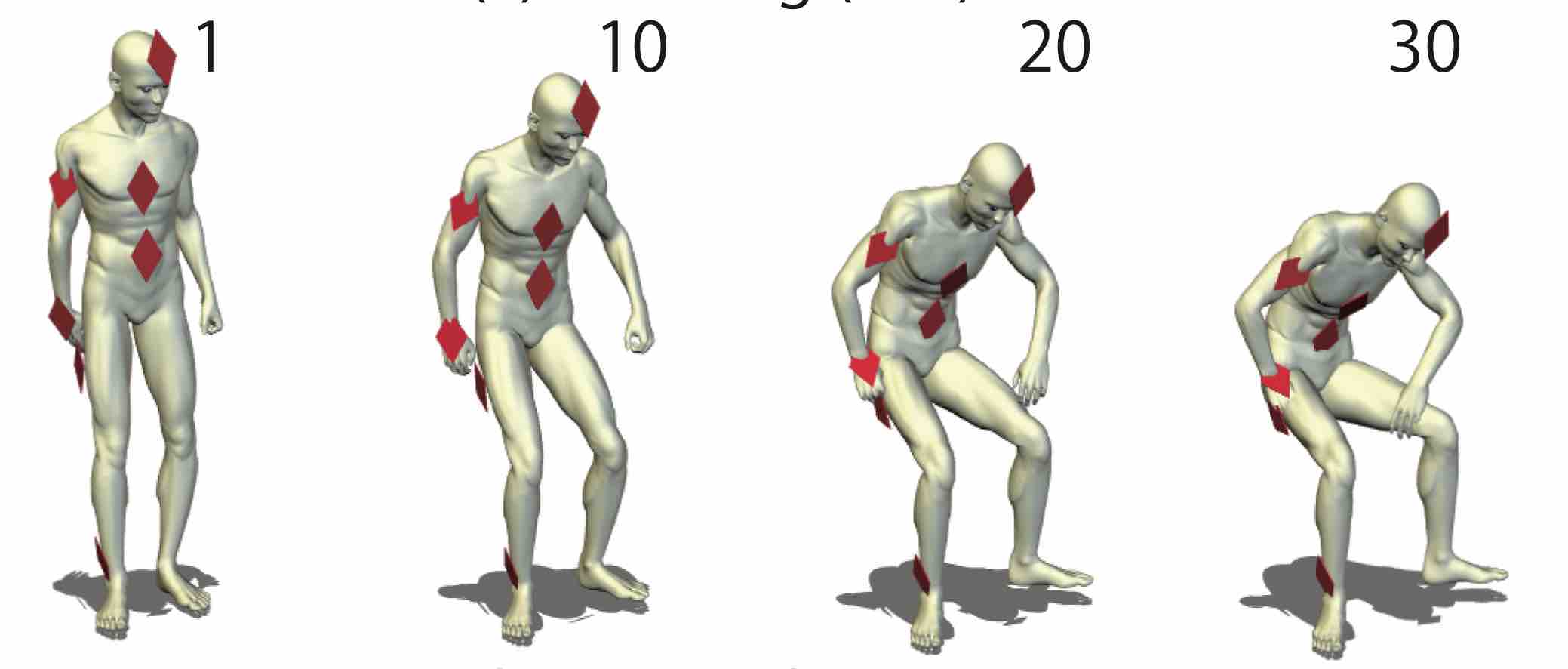}
 \caption{Sitting down}
 \end{subfigure}
 \begin{subfigure}{0.9\columnwidth}
 \centering
 \includegraphics[width=\textwidth,height=2.5cm]{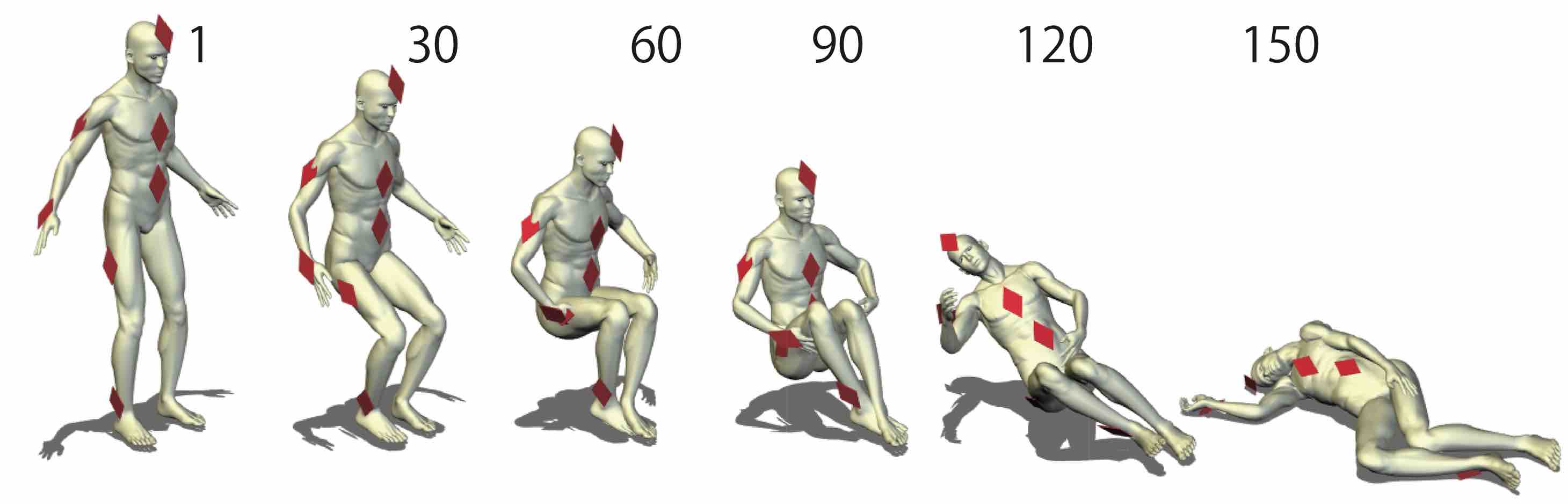}
 \caption{Lying down}
 \end{subfigure}
 \begin{subfigure}{0.9\columnwidth}
 \centering
 \includegraphics[width=\textwidth,height=2.5cm]{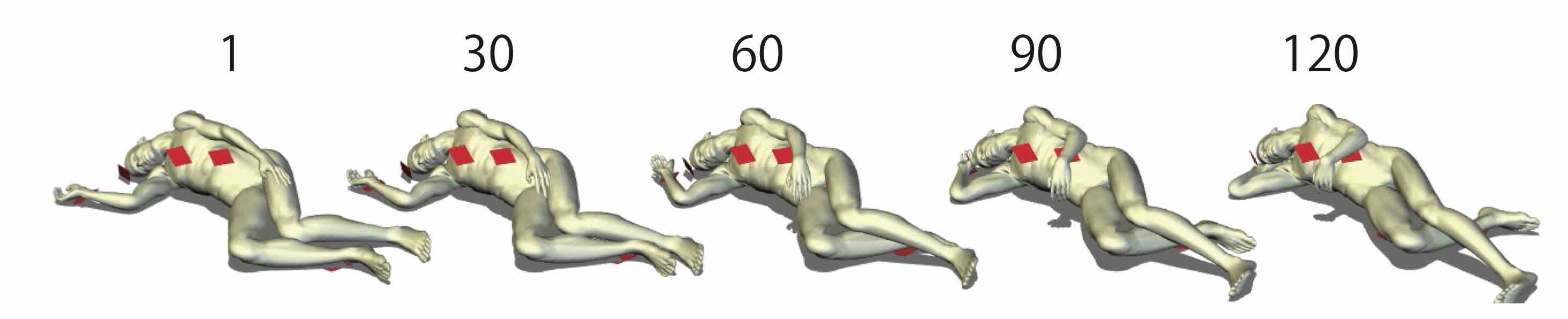}
 \caption{Sleeping}
 \end{subfigure}
 \begin{subfigure}{0.9\columnwidth}
 \centering
 \includegraphics[width=\textwidth,height=2.5cm]{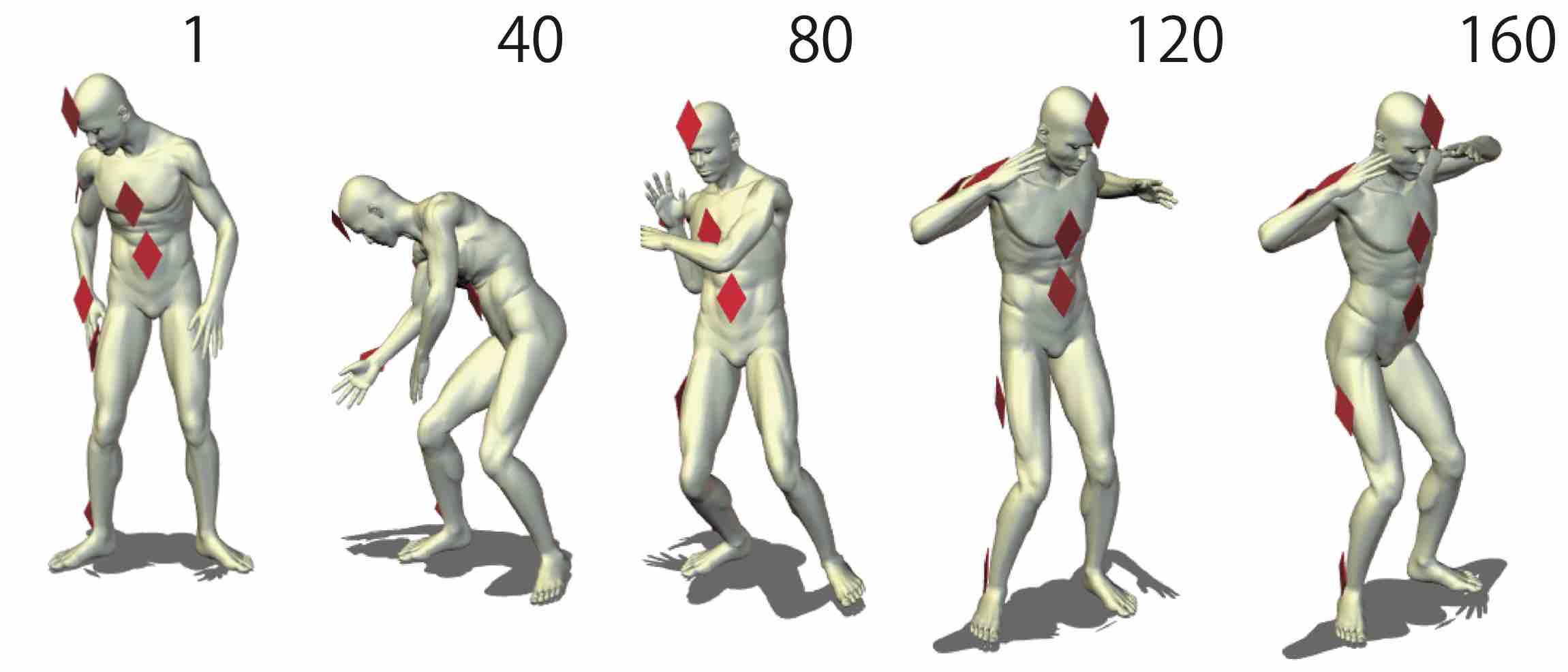}
 \caption{Wearing a jacket}
 \end{subfigure}
 \caption{Postures used in~\cite{channel} to model the WBAN channel (Pictures source: \cite{channel})}
 \label{fig:postures}
\end{figure}

Nodes positions are calculated in $7$ postures: walking (walk), walking weakly (weak), running (run), sitting down (sit), wearing a jacket (wear), sleeping (sleep), and lying down (lie). These postures are represented on Fig. \ref{fig:postures}. 
Walk, weak, and run are variations of walking motions. Sit and lie are variations of up-and-down movement. Wear and sleep are relatively irregular postures and movements.

Channel attenuations are calculated between each couple of nodes for each of these positions as the average attenuation (in dB) and the standard deviation (in dBm). The model takes into account: the shadowing, reflection, diffraction, and scattering by body parts.

\section{CLPB: New Cross Layer Broadcast Protocol} 
\label{NewCrossProtocol}
In this section, we introduce our new cross layer broadcast protocol \emph{CLPB}.
% for \textbf{Cross Layer Protocol for Broadcast} with the particularity of handling broadcast in WBAN. 
%\emph{CLPB} features are to handle \emph{Sink} traffic requirement and to adapt to human body mobility. 
\emph{CLPB} is a slotted protocol that builds on top of pruned communication graphs constructed based of the channel model \cite{channel} . \emph{CLPB} handles both the control medium access and the broadcast process.
%A node \emph{Sink} broadcasts different packets in the network addressed to all nodes along different body part. 
%The ultimate goal is to ensure that all nodes receive all the packets in the shortest delay possible and with the least energy consumption. Our protocol takes into account number of packets to be broadcasted and their transmission frequency.
In oder to include the channel model specificities in the broadcast process \emph{CLPB} needs a \emph{preprocessing} phase that is 
handled at the \emph{Sink} level. After this preprocessing phase, \emph{Sink} broadcast packets that will carry both data and control information (e.g. slots assignment, synchronization information).

\subsection{Preprocessing}
\label{ParagraphTree}
%At the first packet to be sent, \emph{Sink} node triggers slot allocation process. The final allocation result will be followed for subsequent packets to be sent.
The aim of preprocessing is to identify for each posture and for each node, $v$ one or more reliable paths from \emph{Sink} to $v$. This phase is executed only by the \emph{Sink} before starting the broadcast process.
%\begin{enumerate}
%\item Graphs Construction: 

%First, based on our mobility model, 7 communication graphs are constructed (one per posture). 
 
%The mobility model we are using gives us signal attenuation between each couple of nodes for different postures as the average attenuation (in dB) and the standard deviation (in dBm). The idea is to involve channel characteristics in our protocol design: \begin{itemize}
%\item 
First, \emph{Sink} computes, based on the mean and standard deviation of each link in \cite{channel}, the Cumulative Distribution Function (CDF) of the random attenuation $x$ : $F(X)=P[x<X]$ where $X$ is a threshold. $X$ represents the maximum acceptable attenuation referring to the transmission power $-55\,dBm$ and reception sensibility $-100\,dBm$. Thereby, $X$ is equal ($-55$ - ($-100$))=$45\,dBm$ and $F(45)$ represents the probability of a successful transmission. 
%A similar approach is used by authors in \cite{Gewu}.
Then \emph{Sink} defines a pruned communication graph. Nodes in this graph are the nodes in the network, the edges correspond to the links 
% where 
%\item Second, after calculating all connections probability of each link for all the postures, 
%it considers 
%only links 
with transmission probability greater than $0.5$. 
%Note that at the end of the preprocessing the obtained communication graphs are not necessary trees.

\begin{figure}[htbp] 
\begin{subfigure}{\columnwidth}
 \centering
\includegraphics[width=0.30\textwidth,height=3.15cm]{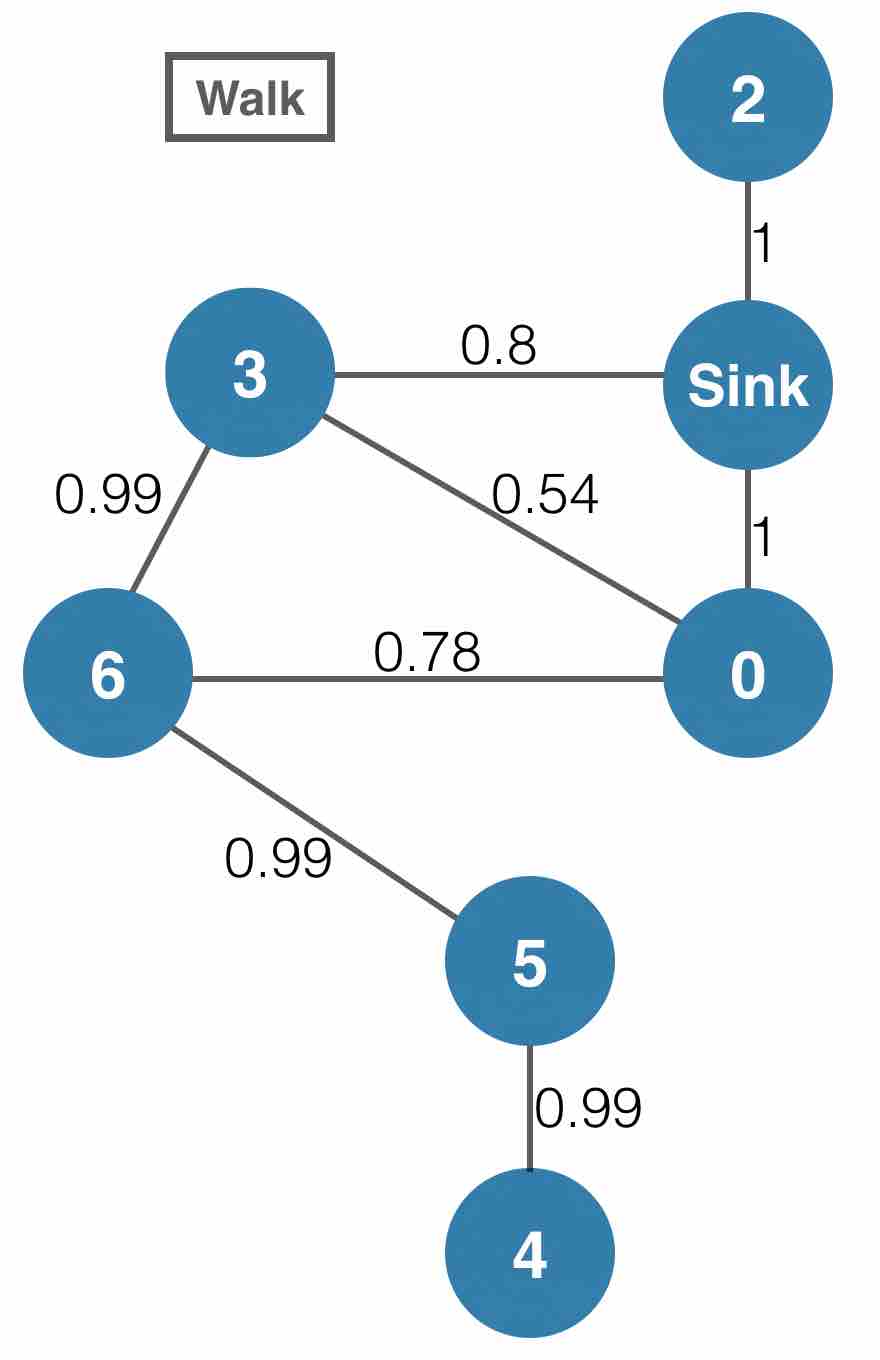}
\includegraphics[width=0.30\textwidth,height=3.15cm]{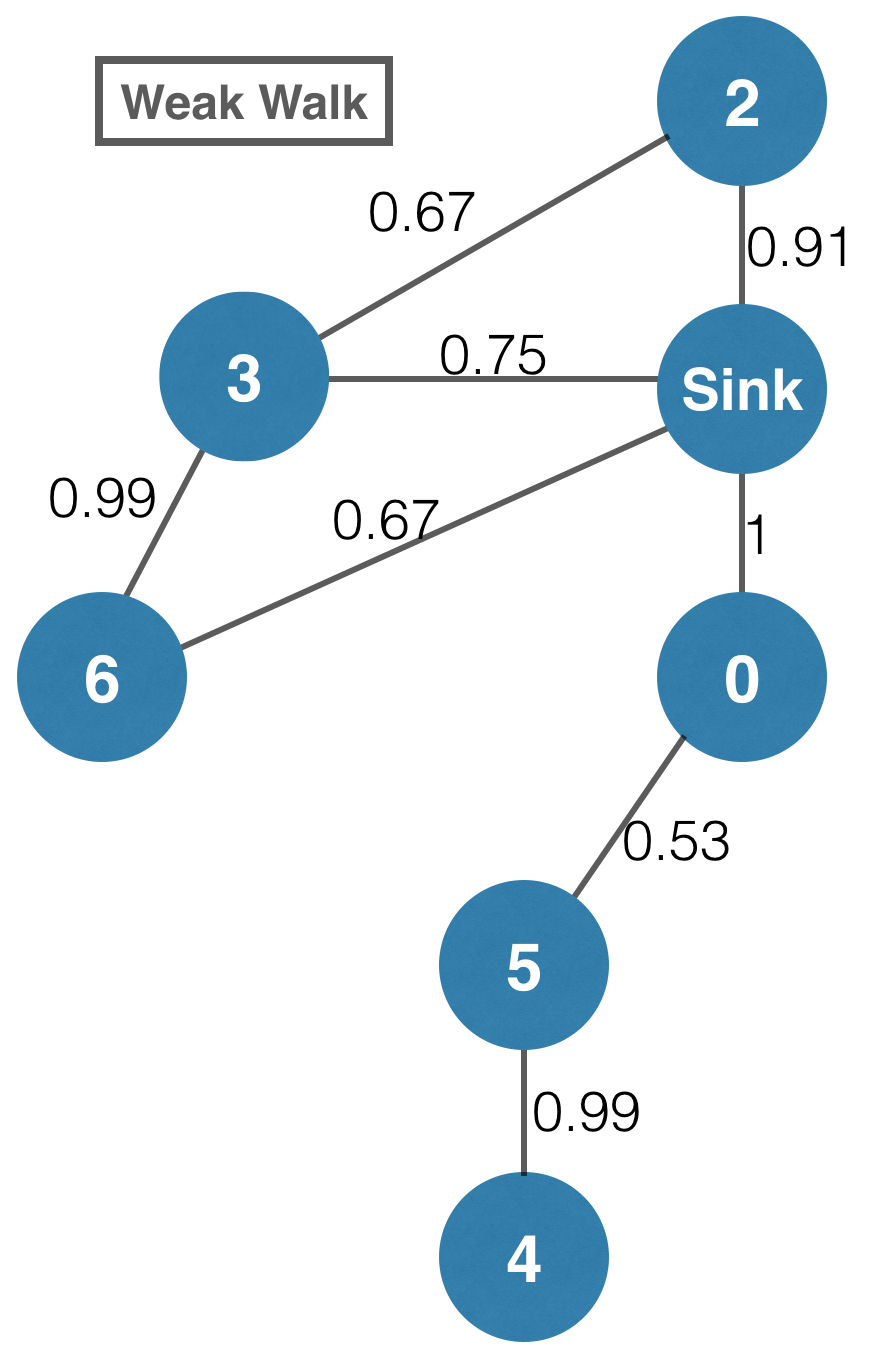}
\includegraphics[width=0.30\textwidth,height=3.15cm]{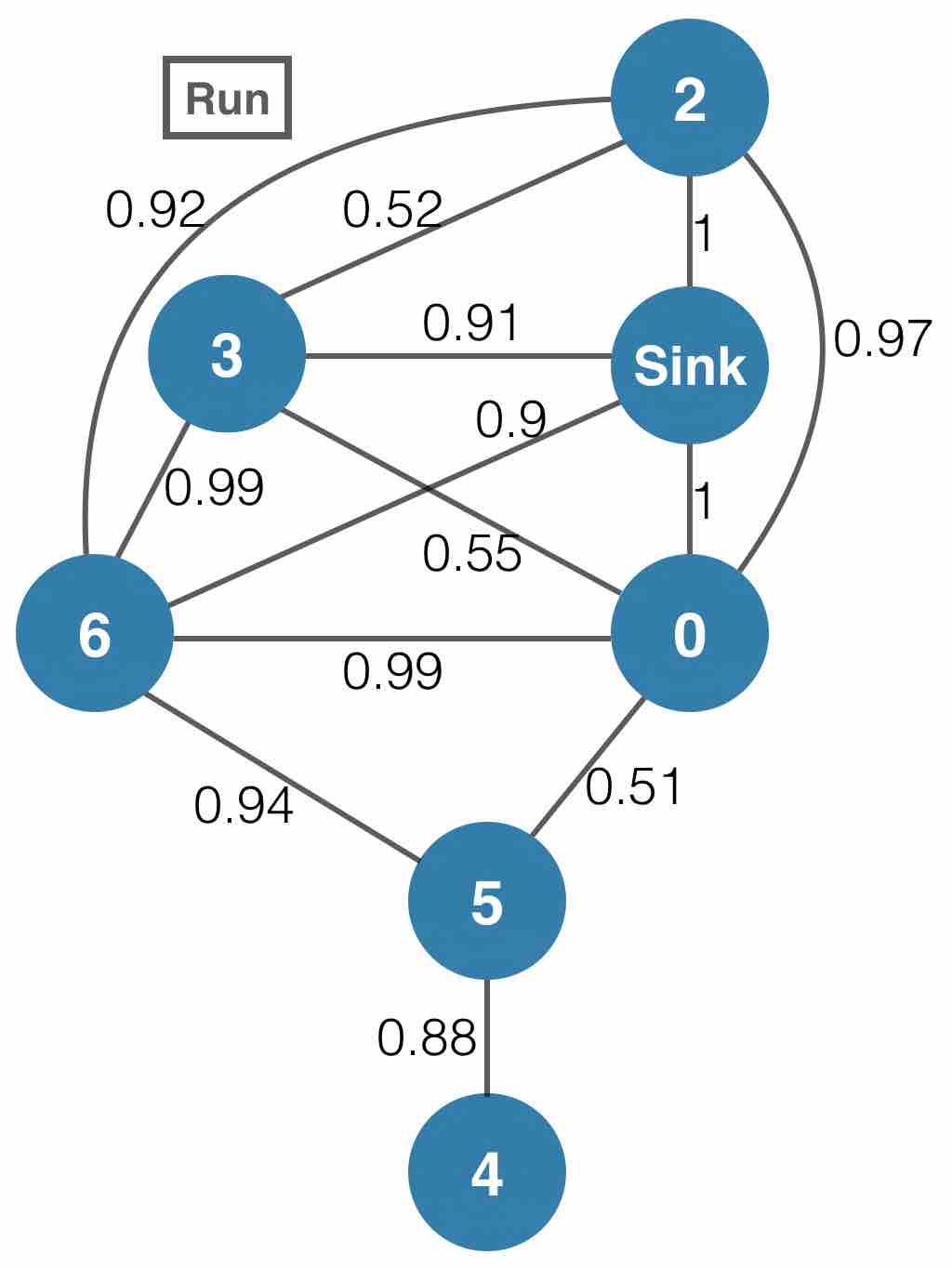}
\end{subfigure}
\begin{subfigure}{\columnwidth}
 \centering
\includegraphics[width=0.24\textwidth,height=3.15cm]{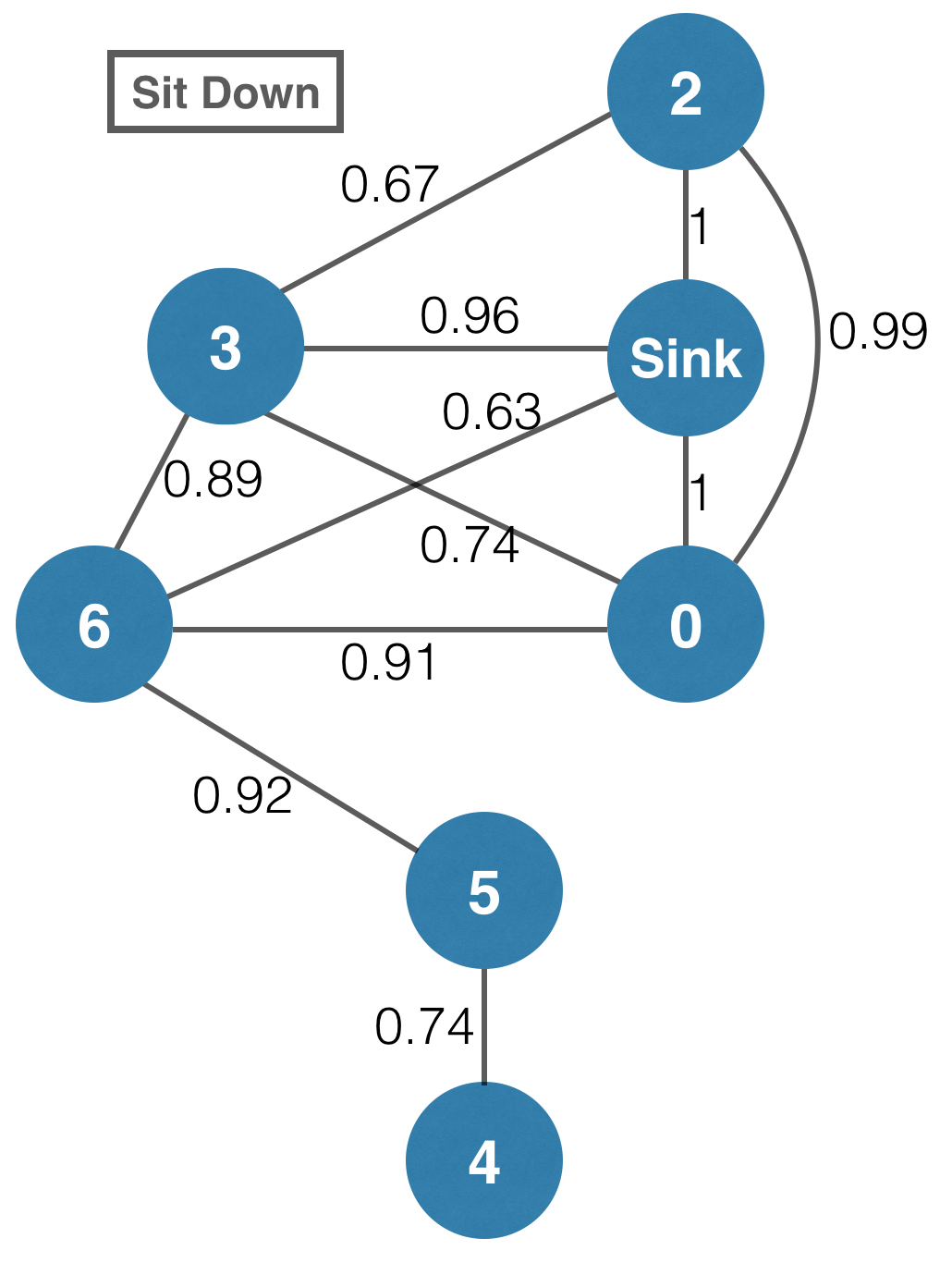}
\includegraphics[width=0.24\textwidth,height=3.15cm]{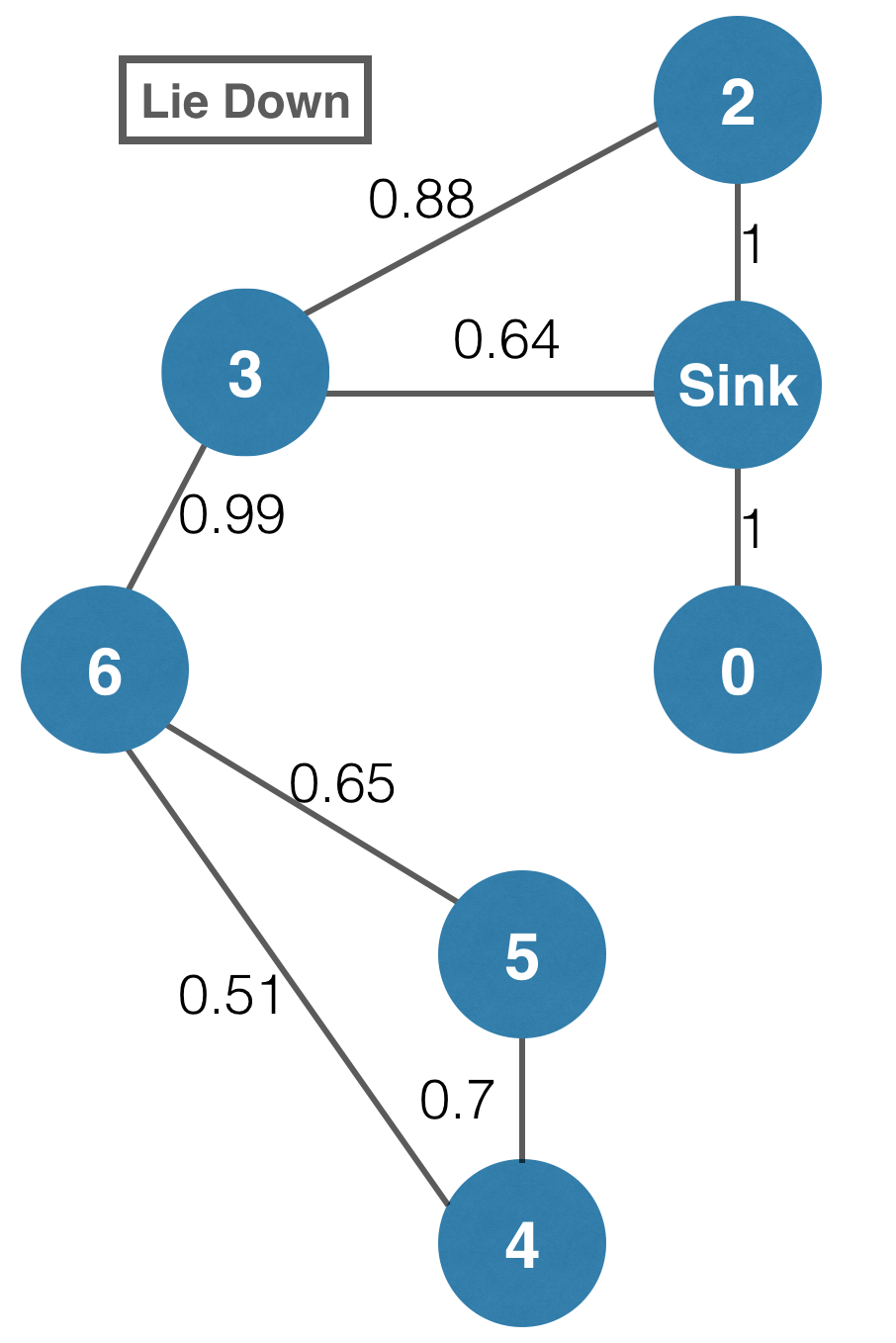}
\includegraphics[width=0.24\textwidth,height=3.15cm]{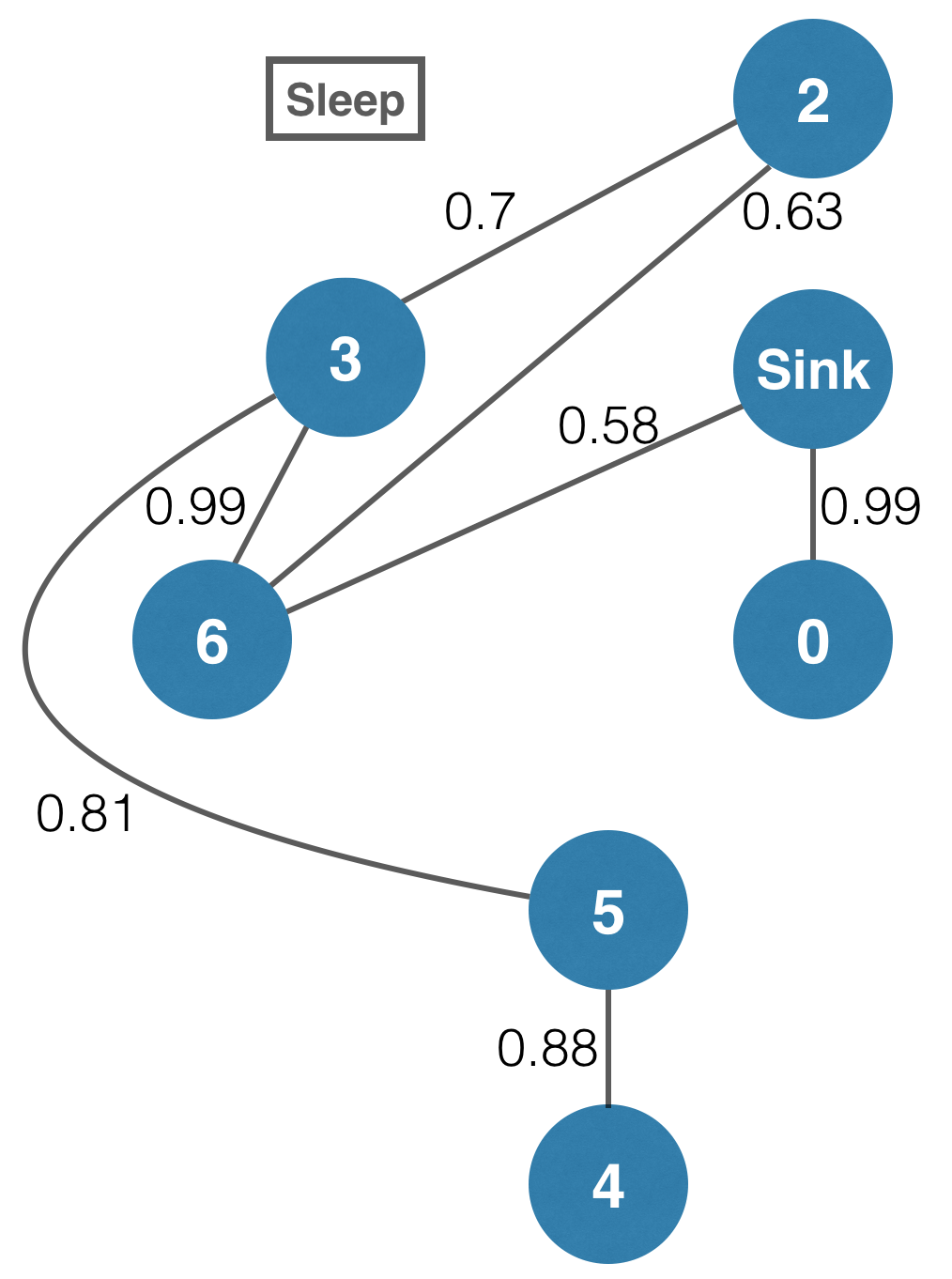}
\includegraphics[width=0.24\textwidth,height=3.15cm]{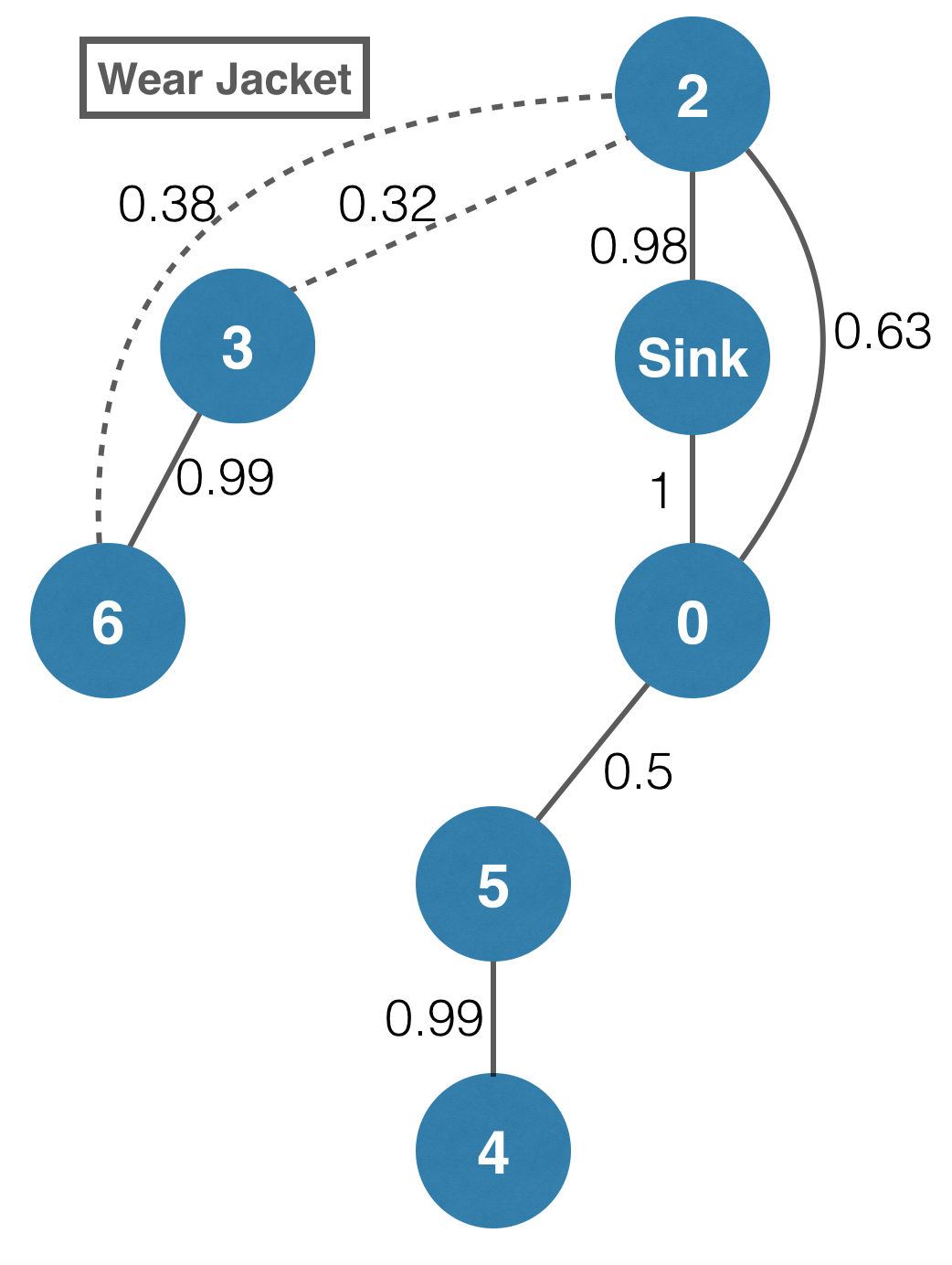}
\end{subfigure}
\caption{Communication Graphs per posture}
 \label{Tree}
\end{figure}

%\item Finally, it iterates until, starting from \emph{Sink} node to any node, all links together represent the highest probability. This aims to minimize the number of transmitting nodes especially if some nodes present unreliable links with neighbors. 
Figure \ref{Tree} shows the 7 pruned communication graphs, one per posture, obtained by applying the procedure described above. Note that at the end of the preprocessing the obtained communication graphs are not necessary trees.
 In the following we will denote these communication graphs $G_i$ where $i$ is the number of the corresponding posture.

Then, \emph{Sink} selects a set of \emph{senders} for each posture. 
For each node, $v$, in the communication graph $G_i$, \emph{Sink} computes the paths between \emph{Sink} and $v$ with maximal reliability. The set of nodes on these paths are included in the \emph{senders} set for the graph $G_i$. In Figure \ref{Tree}, for the communication graph corresponding to the walk posture, the senders set contains nodes $0, 3, 5, 6$.

\subsection{Protocol Overview}
%\emph{Sink} node divides time axis divide into time slots and into cycles. 

%\emph{Sink} node sets up 

%More specifically,
%After the preprocessing phase
\emph{Sink} 
%assigns to each \emph{sender} node one slot to forward data. 
assigns transmission slots to the \emph{senders} outputed by the preprocessing phase (see Section \ref{ParagraphTree}). The broadcasted packets will carry both data and control information (e.g. slots assignment, synchronization information). A node is allowed to forward previously received packets only if it is a sender and the current slot was assigned to it.

%Thus, each node is allowed to forward (broadcast) previously received data only on its designated time slot. 

Finally, \emph{Sink} broadcasts packets that include data and a \emph{medium access and synchronization scheme}. 
Note that with our protocol, no control packets exchange is needed. 

%The time slots assignment is centralized but the coordination is distributed.
Our protocol assumes that nodes execute in synchronized time-slots. Furthermore, it is assumed that the boundaries of slots are also synchronized.

\subsection{Medium access and Synchronization scheme}
\label{scheme}
%Data + scheme
The time is divided at the \emph{Sink} level in \textbf{cycles}.
A cycle corresponds to a sequence of time slots equal to the number of senders, in other words, equal to the number of nodes allowed to transmit including \emph{Sink} node. Cycle duration is given by the following equation eq.\ref{CycleDuration}:
\begin{equation}
\label{CycleDuration}
Cycle Duration = Number Of Senders * Slot Duration
\end{equation}

Nodes synchronize with the sink via the scheduling and synchronization scheme described in details in the sequel.
% nodes are able resynchronize even though some packets are lost.
% However resynchronization is established at the beginning of each cycle.
%Our new cross layer approach minimizes coordination overhead, no exchange of control packets because informations added to data packets are used as control informations and enable nodes to know everything about communication and traffic.
That is, 
each received packet is considered as a reference for the current time slot. A \emph{sender} includes in packets it forwards its slot number called \emph{current slot} and the next cycle start in a way that at the reception of this data packet, nodes can position the current time with respect to the current slot and to the next cycle.

\begin{figure}[htbp]
 \begin{subfigure}{0.9\columnwidth}
 \centering
 \includegraphics[width=\textwidth,height=2.5cm]{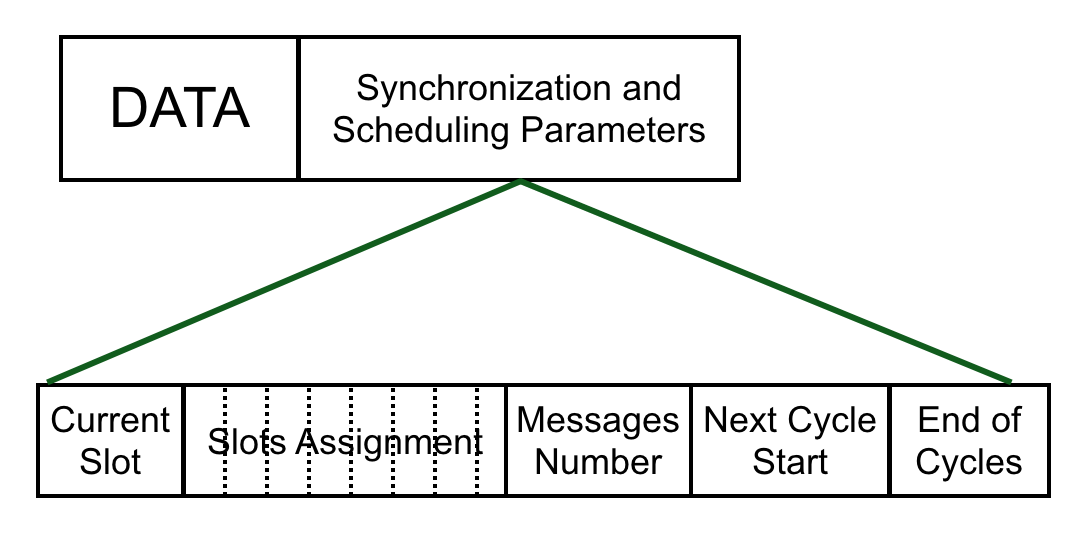}
 \end{subfigure}
 \caption{Packet Description}
 \label{MessageDescription}
\end{figure}

Figure \ref{MessageDescription} presents the synchronization and scheduling parameters. 

\begin{enumerate} 

\item \textbf{Current slot}: is a reference slot that allows nodes to position in time with respect to the current cycle. For example in Figure \ref{ExampleCross}, \emph{Sink} node sends the packet with a current slot equal to its time slot equal to $0$. Suppose, node $3$ misses packet reception from \emph{Sink} node while node $0$ receives it correctly. After that, node $0$ broadcasts the packet with a current slot equal to its time slot (i.e. $2$). Node $3$ receives the packet, it checks slot scheduling, it is a sender but its time slot (equal to $1$) is less than the current time slot (equal to $2$) so it will not broadcast the packet at this cycle and it will schedule the transmission next cycle to avoid collisions.
%cycle or round

\item \textbf{Slots Assignment}: assigns to each sender its time slot. Slot number $0$ is always assigned to \emph{Sink}. Based on example Figure \ref{ExampleCross}, node $3$ sends at time slot $1$ then node $0$ at time slot $2$, while nodes $2$ and $4$ are not designated as senders.

\item \textbf{Messages Number}: represents the total number of packets to be send by \emph{Sink} node and that should be received by all nodes.

\item \textbf{Next Cycle Start}: Depends on the transmission frequency of \emph{Sink} node. This frequency allows to compute the time between two consecutive cycles: the \emph{CyclesInterleave} parameter presented below and so to determine \emph{Next Cycle Start}. If the transmission frequency is such that \emph{Sink} node receives an application packet while previous packet is still in broadcast in the network (the current cycle is not finished), then cycles interleave is nul. As shown in the example \textbf{a)} in Figure \ref{ExampleCycleGap}, \emph{Sink} node receives an application packet every $2$ time slots, which is less than a cycle duration ($5$ time slots). In this case, \emph{Sink} node puts packets in its buffer and waits the end of the current cycle. Then, it immediately starts a new cycle. If \emph{Sink} node receives an application packet much later. For example, as shown in the example \textbf{b)} in Figure \ref{ExampleCycleGap}, \emph{Sink} node receives an application packet every $8$ time slots, which is greater than a cycle duration ($5$ time slots). In this case, nodes enter in a sleep mode waiting for the next cycle.

\begin{align}
Cycles Interleave = \left \{ 
 \begin{array}{l}
 				0 \mbox{, if \emph{Sink} transmission frequency $<$ Cycle Duration} \\
 (\lceil Transmission Frequency / Slot Duration\rceil) \\
		 * Slot Duration] - Cycle Duration (eq. \ref{CycleDuration}) \mbox{, otherwise.}
		 \end{array}\right .
\end{align}

\begin{figure}[htbp]
 \begin{subfigure}{0.9\columnwidth}
 \centering
 \includegraphics[width=\textwidth,height=4.5cm]{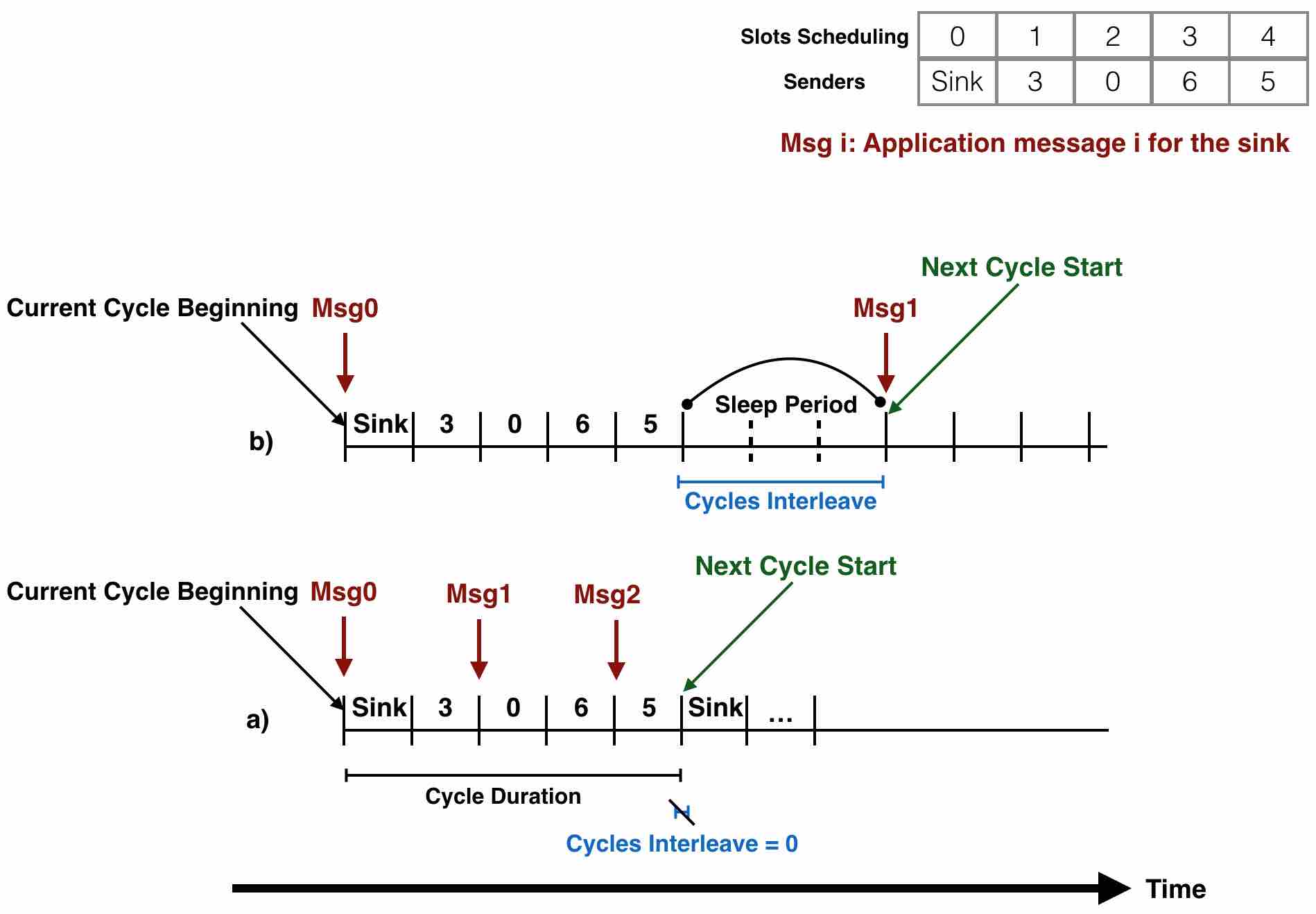}
 \end{subfigure}
 \caption{Cycles Interleave: Both cases}
 \label{ExampleCycleGap}
\end{figure}

 \emph{Next Cycle Start} is a key parameter that optimizes nodes duty cycle. In fact, instead of alternating between reception and sleep mode each time slot, nodes will go back to sleep and schedule wake up when more packets are available in the network.

\item \textbf{\emph{End of Cycles}} indicates to nodes when to sleep definitely. In fact, nodes keep waking up each cycle until receiving all data packets. However, in case of packet loss due to unreliable links for example, nodes will keep waiting for lost packets and continue to wake up each time slot. In order to optimize nodes energy consumption, \emph{Sink} node will compute based on traffic parameter an estimation for time needed to send all packets. Then, if a node reaches the estimated time and is still missing some packets, it will be able to detect this loss and decide to sleep definitely. 

To avoid that nodes miss data packet, it is important to over estimate \emph{End of Cycles} parameter by supposing the worst case given by the Equation \ref{Cycle-end} below. 
%Thus, we suppose that, in worst case, a sender has to delay broadcast to its time slot at the next cycle to send each packet. This gives as:
\begin{equation}
\label{Cycle-end} 
End Of Cycles = Messages Number * Cycle Duration (eq. \ref{CycleDuration}).
\end{equation}

\end{enumerate}
\subsection{Protocol Details Description}

%Figure \ref{ExampleCross} resumes a simple broadcast scenario with our new cross layer protocol. 

 \begin{figure}[htbp]
 \begin{subfigure}{0.9\columnwidth}
 \centering
 \includegraphics[width=\textwidth,height=4.5cm]{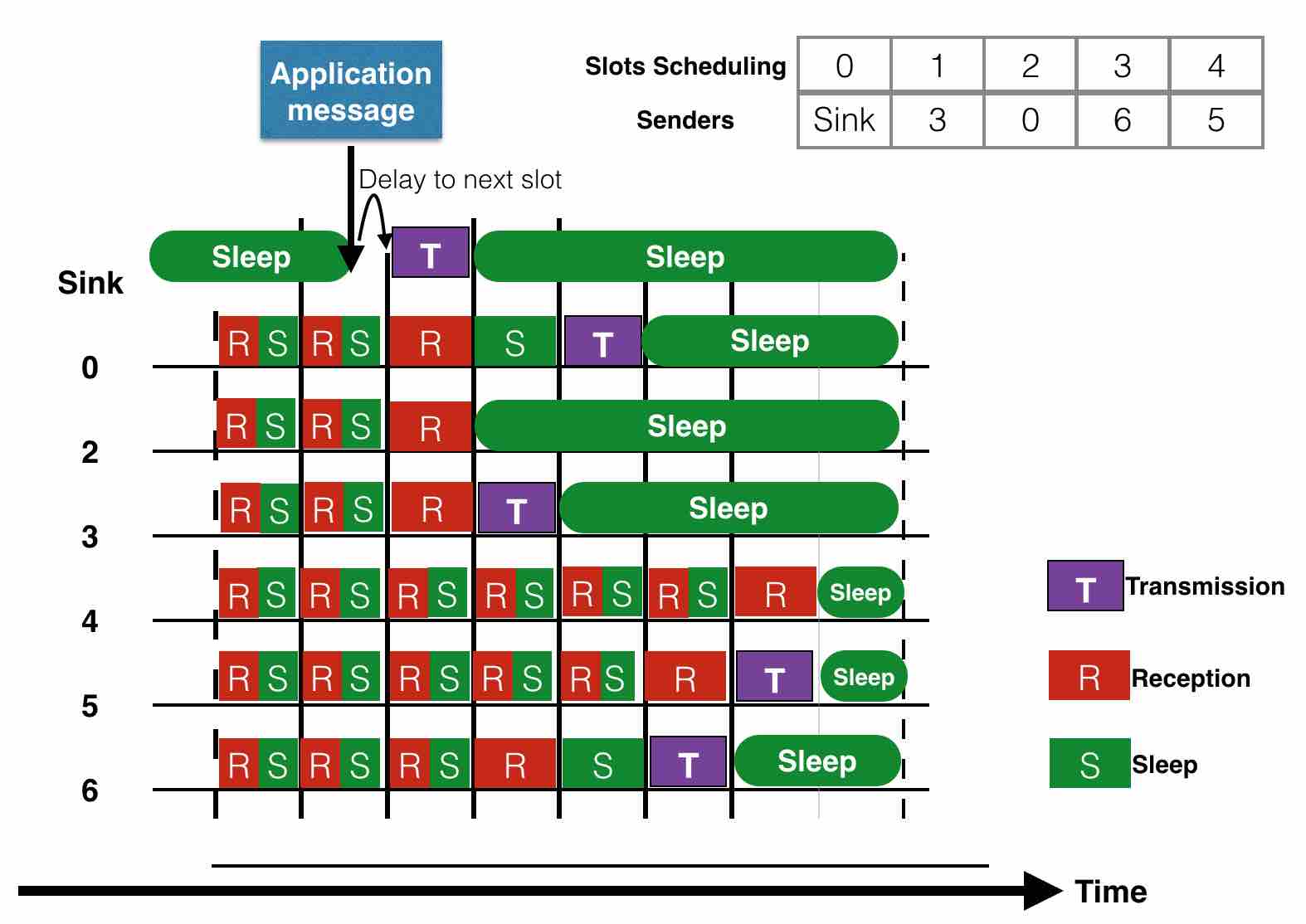}
 \end{subfigure}
 \caption{CLPB execution for posture WALK}
 \label{ExampleCross}
\end{figure}

Upon reception of an application packet \emph{Sink} schedules the broadcast of the packet plus the medium access control scheme information (see Section \ref{scheme}) at the next time slot. 

Each \emph{sender} node wakes up on each slot on reception mode, RX, for a period of time equal to a half time slot. During this period, there are two possible cases: 
\begin{itemize}
\item no packet is received: Node goes back to sleep mode and wakes up (on reception mode RX) next time slot.
\item a packet is received:
 \begin{itemize}
 \item if it is allowed to transmit (based on time slots schedule), then it schedules a transmission.
 \item if more packets are expected (based on sink transmission frequency) then it computes the next cycle and goes back to sleep at the end of the current cycle. Otherwise, it sleeps definitely. 
 \end{itemize}
\end{itemize}

When a node wakes up on its assigned time slot to transmit previously received data, it transmits until the end of its slot.
% data packets in the buffer while there are still packets waiting in buffer and while it stills on its time slot. 
After this point, it delays the remaining packets to be broadcasted in the next cycle.
%different cases and different traffic param 
 
Nodes alternate between: Reception, Sleep and Transmission modes. However, differently from other similar techniques, we strive to reduce the number of state switches and the duty cycle duration. As shown in Figure \ref{ExampleCross} for example, after receiving a packet, nodes $0$ and $6$ sleep waiting their turn to broadcast based on the scheduling scheme. Also, node $3$ sleeps after broadcasting the packet. Leaf nodes (in this example: nodes $2$ and $4$) sleep after receiving all packets while sender nodes sleep after transmitting all received packets. Note that in this example, we suppose only one packet is broadcasted in the network so nodes sleep after receiving the unique packet. If more than one packet is expected, nodes sleep at the end of the current cycle and wake up the next cycle.

\section{Performance analysis}
\label{sec:simulation}
%%%%%%%%%%%%%%%%%%%%%%%%%% Broadcast strategies
In this section we compare the strategies proposed in \cite{BCPP15} (and its companion technical report \cite{badreddine:hal-01404848}) and our new cross layer protocol face to various broadcast rates up to 1000 packets/s and seven realistic body posture and mobility.
We use the discrete event simulator Omnet++ \cite{OmnetHomePage} and the Mixim framework ~\cite{mixim} enriched with the channel and mobility model \ref{ChannelModel}.
Broadcast strategies investigated in \cite{BCPP15} and its companion technical report \cite{badreddine:hal-01404848} are as follows: \textit{Flooding} represents the basic broadcast strategies where nodes rebroadcast each received packet; \textit{Plain flooding} is a more restrictive strategy where nodes rebroadcast each received packet only once; \textit{Pruned flooding} strategy is based on a random choice of the next hops; \textit{Probabilistic flooding (P=P/2)} where broadcasting decision depends on a probability $P$. $P$ is divided by $2$ after every broadcast so that broadcasting is more limited by time; \textit{MBP} that begins as a basic flooding algorithm (i.e \emph{Flooding} strategy) then it is conditioned by different parameters. The aim is to limit broadcasting at the center of the network while redirecting it to the peripheral nodes; \emph{OptFlood} strategy for Optimized Flooding. It is presented as a revised version of \emph{Flooding}. This strategy targets the good end-to-end delay similar to \emph{Flooding} while lowering energy consumption by eliminating unnecessary retransmissions. 

\subsection{Simulation Settings}
\label{SimulationSettings}
Above the channel model described in section \ref{ChannelModel}, we used standard protocol implementations provided by the Mixim framework \cite{mixim}. In particular, we used, for the medium access control layer, the IEEE 802.15.4 implementation. The sensitivity levels, header length of the packets and other basic information and parameters are taken from the 802.15.4 standards.

Each data point is the average of $50$ simulations run with different seeds. The transmission power is set at the minimum limit level $-55\,dBm$ that allows an intermittent communication given the channel attenuation and the receiver sensitivity $-100\,dBm$, guarantees a connected network at each time t of the simulation and ensures a limited energy consumption. 
Slot duration is equal to $5ms$ with a bitrate equal to $1Mb/s$.

Our evaluation targets the parameters below:
\begin{itemize}
 \item \textbf{Percentage of covered nodes:} Since our unique source is the \emph{Sink}, we therefore calculate the percentage of nodes that have received a packet.
 \item \textbf{Percentage of de-sequencing}: The percentage of packets received in a different order than the sending order. This parameter is evaluated only in the case of broadcast rates greater than 1. 
 \item \textbf{End to end delay:} The average end-to-end delay is the time a packet takes to reach the destination(i.e. every node except the Sink).
 \item \textbf{Energy Consumption:} This is a main concern in WBAN. Energy consumption is estimated as \emph{the number of transmissions and receptions} in the case of single packet broadcast and as \emph{average number of transmissions and receptions per node} in the case of higher than 1 broadcast rates.
\end{itemize}.
 
Sections \ref{CLPBperf} and \ref{BufferSize300} below present simulation results when strategies are stressed with broadcast rates from 2 to 1000 packets/s and various buffer sizes (e.g. 100, 200, 300 etc) \footnote{Due to lack of space we choose to present the results for buffer sizes 100 and 300.}. 
The goal of studying strategies performance with various transmission rates and different buffer sizes is to highlight the hidden impact of some parameters like MAC buffer size on strategies performances. Our simulation confirm that cross-layer approach offers the best performances. 
Section \ref{NewVsOld} zoom the case when \emph{Sink} node broadcasts only $1$ packet. The results confirm that even for small rates broadcast the cross-layer approach offers the best performances.

%%%%%%%%%%%%%%%%%%%%%%%%%%%%%%%%%%%%%%%%%%%%%%%%%%%%%%%%%%%%%%%%%%%%%%%%%%%%%%%%%%%%%%%%%%%%%%%
\subsection{Broadcast rates up to 1000 packets/s and buffer size 100}
\label{CLPBperf}
We stress \emph{CLPB} and the flat strategies in \cite{BCPP15} and \cite{badreddine:hal-01404848} with broadcast rates up 1000 packets/s. 
In all the seven body postures \emph{CLPB} strategy outperforms the flat strategies. Moreover, the good performances \emph{CLPB} are maintained for broadcast rates up to 500 packets/s while the strategies in \cite{BCPP15} observe a drop of performances starting with 10 packets/s. 
 
\subsubsection{Percentage of covered Nodes}

\begin{figure}[htbp]
\centering
 \begin{subfigure}{0.8\columnwidth}
 \centering
 \includegraphics[width=\textwidth,height=2.95cm]{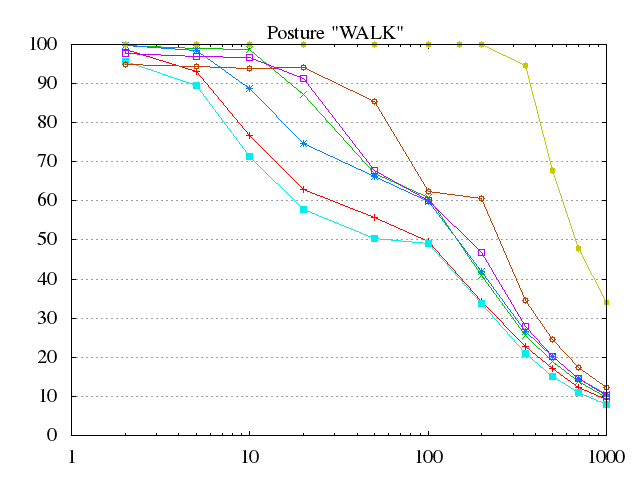}
 \end{subfigure}
 \begin{subfigure}{0.8\columnwidth}
 \centering
 \includegraphics[width=\textwidth,height=2.95cm]{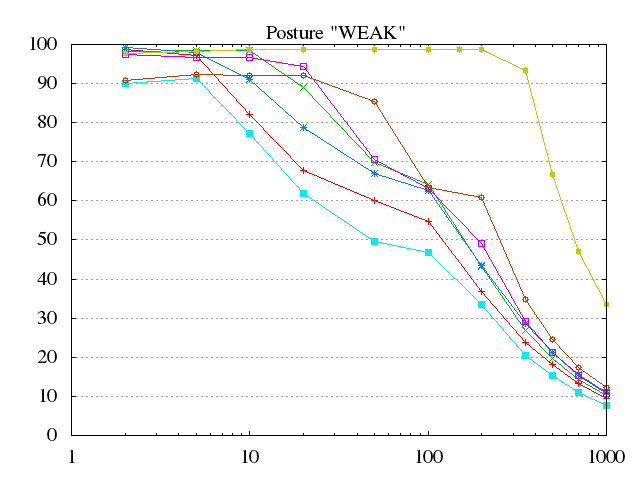}
 \end{subfigure}
 \begin{subfigure}{0.8\columnwidth}
 \centering
 \includegraphics[width=\textwidth,height=2.95cm]{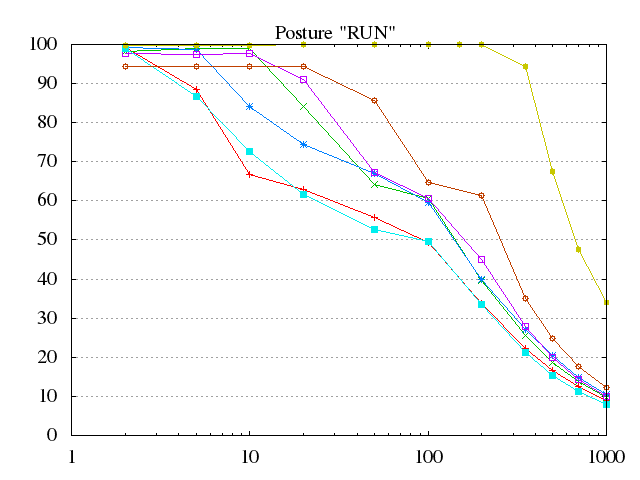}
 \end{subfigure}
 \begin{subfigure}{0.8\columnwidth}
 \centering
 \includegraphics[width=\textwidth,height=2.95cm]{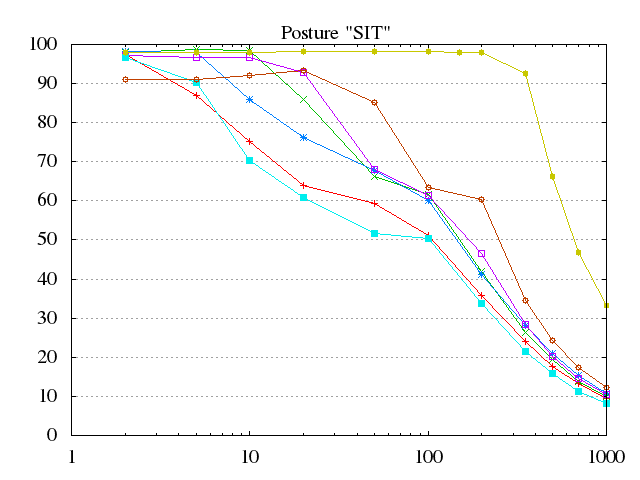}
 \end{subfigure}
 \begin{subfigure}{0.8\columnwidth}
 \centering
 \includegraphics[width=\textwidth,height=2.95cm]{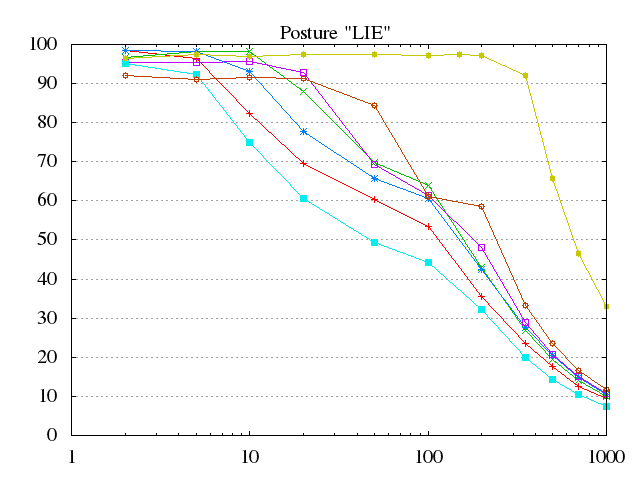}
 \end{subfigure}
 \begin{subfigure}{0.8\columnwidth}
 \centering
 \includegraphics[width=\textwidth,height=2.95cm]{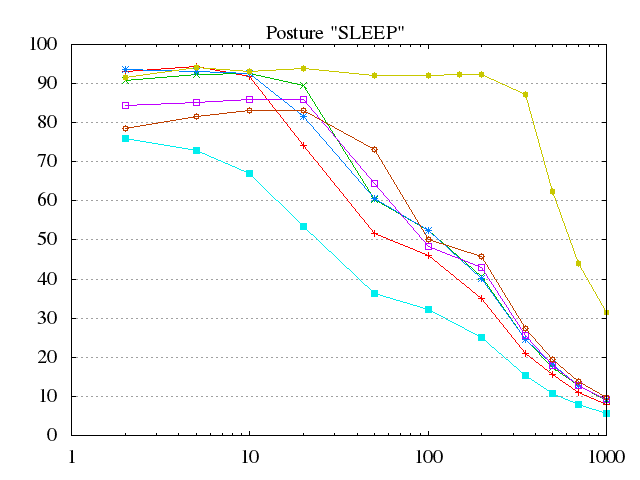}
 \end{subfigure}
 \begin{subfigure}{0.8\columnwidth}
 \centering
 \includegraphics[width=\textwidth,height=4.35cm]{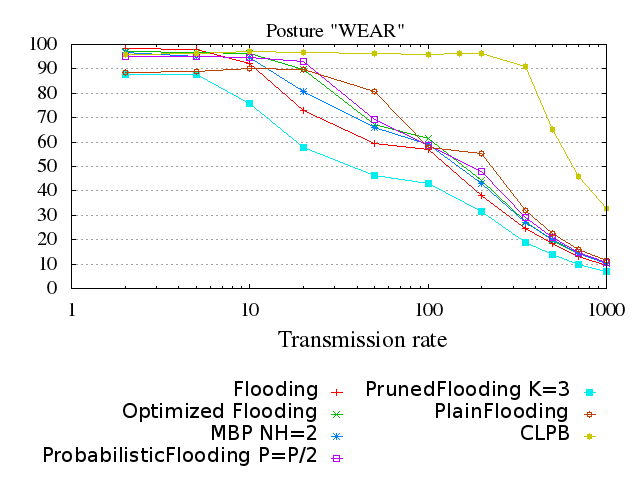}
 \end{subfigure}
 \caption{Percentage of covered nodes in function of transmission rate for all postures }
\label{PercentStressAll100}
\end{figure}

Figure \ref{PercentStressAll100} presents the percentage of covered nodes function of \emph{Sink} transmission rate for all postures. This rate is presented as the number of packets per second.

All broadcast strategies behave similarly: Going to $1000$ packets/s, the percentage of covered nodes almost linearly decreases to reach $10$\%. At $100$ packets/s, percentage of covered nodes is, in average, equal to $50$\%.

%However, at the beginning, strategies manage to maintain an almost constant percentage. With a low transmission rate, strategies can preserve a high percentage of covered nodes. 

With \emph{Flooding} strategy, percentage decreases starting from $5$ packets/s. Nodes broadcast without restrictions received packets which overloads the network, creates collisions and so packets loss. \emph{Pruned Flooding} and \emph{MBP} percentage decreases starting from $5$ packets/s too. With \emph{Pruned Flooding}, even if nodes are restricted to broadcast to only $K$ nodes, still, with $K=3$ many packets copies are generated and the network is overloaded. With \emph{MBP}, broadcast is only delayed to give time for other nodes to receive and acknowledge correct reception. This technique allows \emph{MBP} to give better percentage than \emph{Flooding} and \emph{Pruned Flooding} strategies because it limits collisions. In SLEEP posture, \emph{Flooding} and \emph{MBP} strategies are able to maintain a good percentage of covered nodes equal to $91\%$ up to $10$ packets/s. Due to low mobility and less available links, network is less overloaded so less collisions and less packets loss. However, in more mobile and dense postures, performance decreases, for example, in RUN posture, \emph{Flooding} strategy shows $66\%$ of covered nodes up to $10$ packets/s and $75\%$ in SIT posture. 

\emph{PrunedFlooding} presents the lowest percentage. It is due to the important amount of packets generated in the network which creates collisions and packets loss. And also, for the random choice of next hops, some nodes are not qualified for forwarding. The results of the other strategies are close at the beginning then overlap and converge to the same point.

\emph{CLPB} maintains a good percentage, greater than $90$\%, up to $350$ packets/s. With $350$ packets/s, \emph{Sink} has one packet to send each $0.00285$s. Suppose a cycle lasts $5$ time slots with a time slot duration equal to $0.005$s. At the end of the cycle, \emph{Sink} node has $8$ packets waiting in buffer for broadcast. Or, with a bit rate equal to $1Mbs$, \emph{Sink} node can send up to $5$Kbs during its time slot. A packet size is equal to $544$ bits then \emph{Sink} node can send up to ($5$Kbs / $544$ bits) packets equal to $9$ packets per time slot.
Beyond $350$ packets/s, performance falls to $30$\% of covered nodes by $1000$ packets/s. All nodes are enable to broadcast all waiting packets and empty the buffer. 
%As for the other broadcast strategies, buffer size impacts strategy performance. For this reason, simulation results will be presented for a different buffer size in section \ref{BufferSize300}.

% \begin{figure}[htbp]
% \begin{subfigure}{0.9\columnwidth}
% \centering
% \includegraphics[width=\textwidth,height=5.25cm]{Pictures/NewCrossLayer/CLPBstress/FigPosturesPercentWafa.jpg}
% \end{subfigure}
% \caption{Percentage of covered nodes for all postures }
% \label{ReceivedMsgCrossStress}
%\end{figure}

\subsubsection{Percentage of De-sequencing}
\label{Deseq100}
\begin{figure}[htbp]
\centering
 \begin{subfigure}{0.8\columnwidth}
 \centering
 \includegraphics[width=\textwidth,height=3cm]{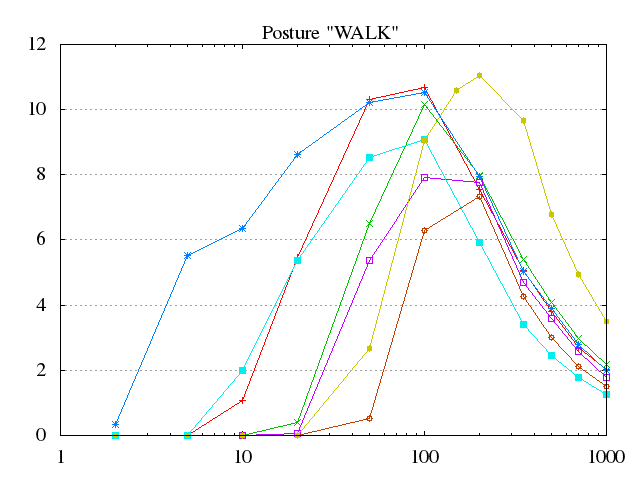}
 \end{subfigure}
 \begin{subfigure}{0.8\columnwidth}
 \centering
 \includegraphics[width=\textwidth,height=3cm]{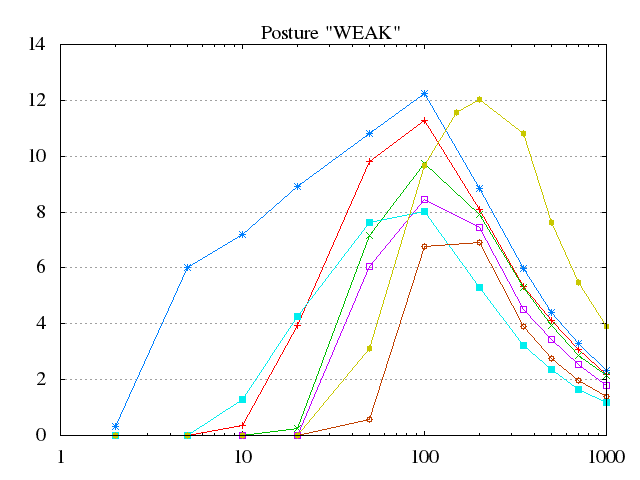}
 \end{subfigure}
 \begin{subfigure}{0.8\columnwidth}
 \centering
 \includegraphics[width=\textwidth,height=3cm]{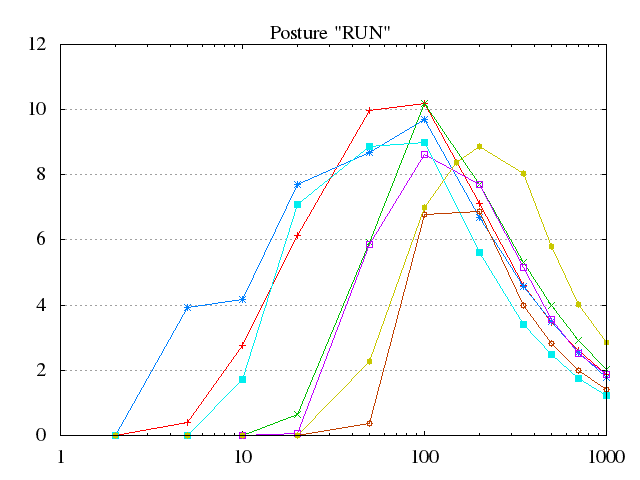}
 \end{subfigure}
 \begin{subfigure}{0.8\columnwidth}
 \centering
 \includegraphics[width=\textwidth,height=3cm]{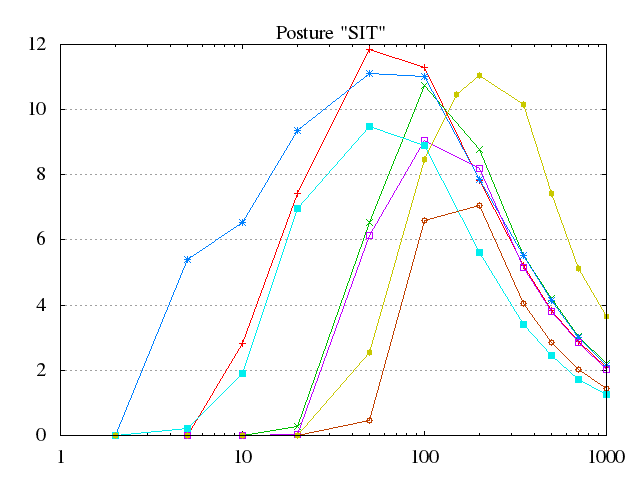}
 \end{subfigure}
 \begin{subfigure}{0.8\columnwidth}
 \centering
 \includegraphics[width=\textwidth,height=3cm]{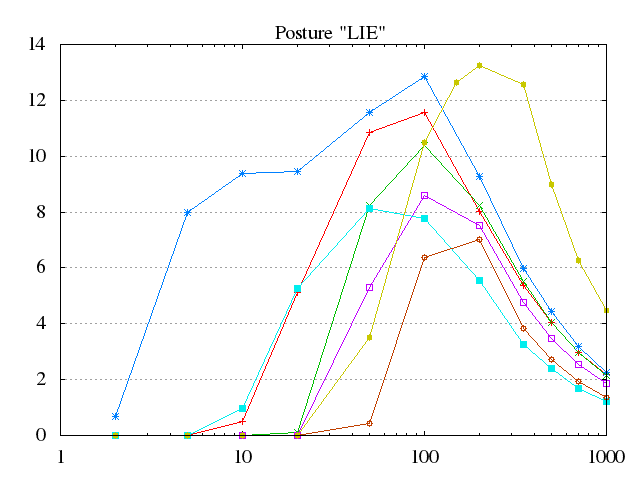}
 \end{subfigure}
 \begin{subfigure}{0.8\columnwidth}
 \centering
 \includegraphics[width=\textwidth,height=3cm]{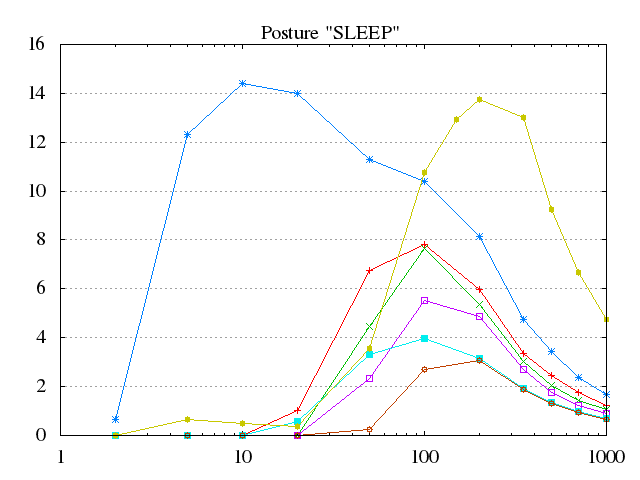}
 \end{subfigure}
 \begin{subfigure}{0.8\columnwidth}
 \centering
 \includegraphics[width=\textwidth,height=4.35cm]{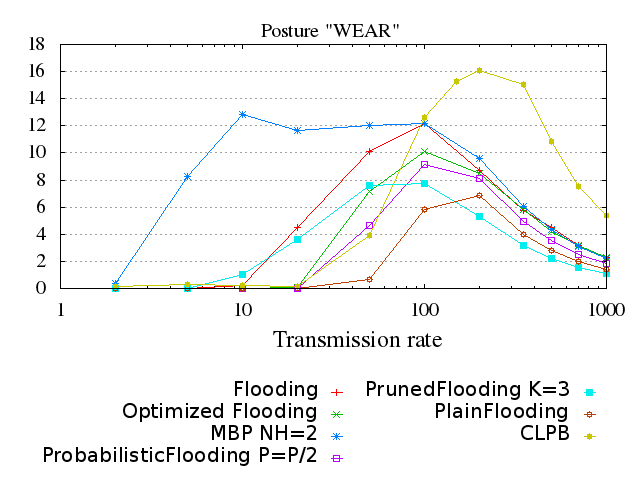}
 \end{subfigure}
 \caption{Percentage of De-sequencing in function of transmission rate for all postures }
\label{DeseqStressAll100}
\end{figure}

Figure \ref{DeseqStressAll100} presents the percentage of de-sequencing for all postures in function of transmission rate.

Three phases can be observed:
\begin{itemize}
\item At the beginning, all strategies present $0$\% of de-sequencing. At this point, strategies are able to handle more than one packet in the network.
\item Then, from a given rate (depending on the strategy), the percentage increases. Here, based on Figure \ref{PercentStressAll100}, the percentage of covered nodes decreases due to collisions and packets loss. Therefore, sequencing is no longer ensured.
\item Finally, the percentage decreases to converge to 0\% again due to the fact that few packets are received.
\end{itemize}

\emph{MBP} strategy presents the highest percentage of de-sequencing for all postures, starting from $2$ packets/s. Nodes can broadcast immediately or delay received packet broadcasting. To take the decision, they compare each packet number of hops to a threshold \emph{NH}. Thus, de-sequencing is more feasible.
 
Percentage of de-sequencing increases starting from $5$ packets/s for \emph{Flooding} and \emph{Pruned Flooding} and from $20$ packets/s for \emph{Optimized Flooding}, \emph{PlainFlooding} and \emph{Probabilistic Flooding} for most postures. \emph{Flooding} and \emph{Pruned Flooding} have difficulties to handle transmission rate increase due to collisions and packets loss. An exception with SLEEP posture, where the percentage of de-sequencing is observed starting from $10$ packets/s due to the nature of the posture where few links are reliable. 

%\emph{PlainFlooding} low percentage of de-sequencing is due to a low percentage of covered nodes.

\emph{Optimized Flooding} strategy presents the lowest percentage of de-sequencing compared to \emph{Flooding} and \emph{MBP} strategies. 

\emph{CLPB} reacts as the other strategies and we observe a bit de-sequencing. This is due to the mobility model. That is, unreliable links may occur, thus allowing reception of one of several packets. The links then disappear and the complete sequence will be received through a more reliable and legitimate link. 

%\emph{Due to lack of space, the results presented in the sequel focus only postures: WALK, RUN and SLEEP. The results for the other postures are similar.}

\subsubsection{Average number of transmissions and receptions per node}
\begin{figure}[htbp]
\centering
 \begin{subfigure}{0.8\columnwidth}
 \includegraphics[width=\textwidth,height=3cm]{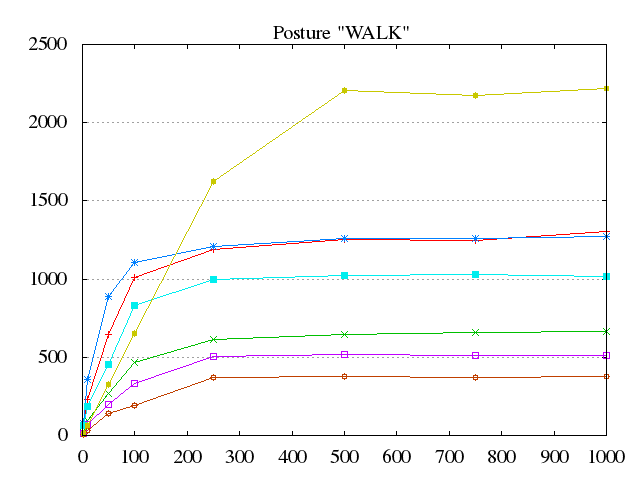}
 \end{subfigure}
 \begin{subfigure}{0.8\columnwidth}
 \includegraphics[width=\textwidth,height=3cm]{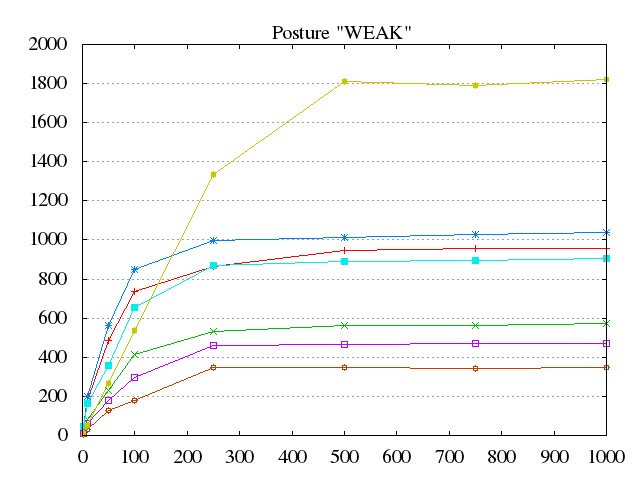}
 \end{subfigure}
 \begin{subfigure}{0.8\columnwidth}
 \centering
 \includegraphics[width=\textwidth,height=3cm]{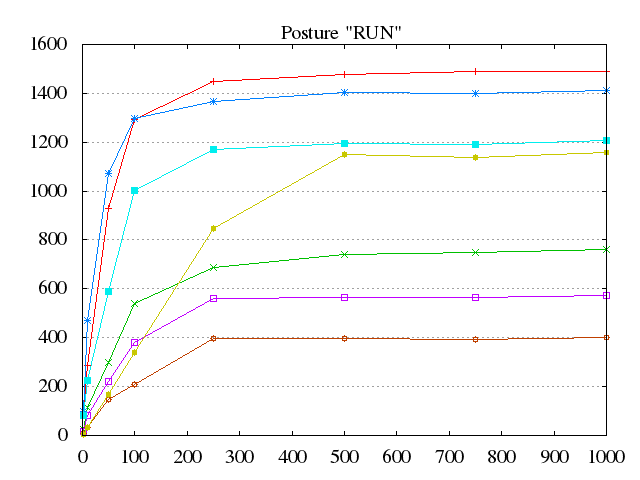}
 \end{subfigure}
 \begin{subfigure}{0.8\columnwidth}
 \centering
 \includegraphics[width=\textwidth,height=3cm]{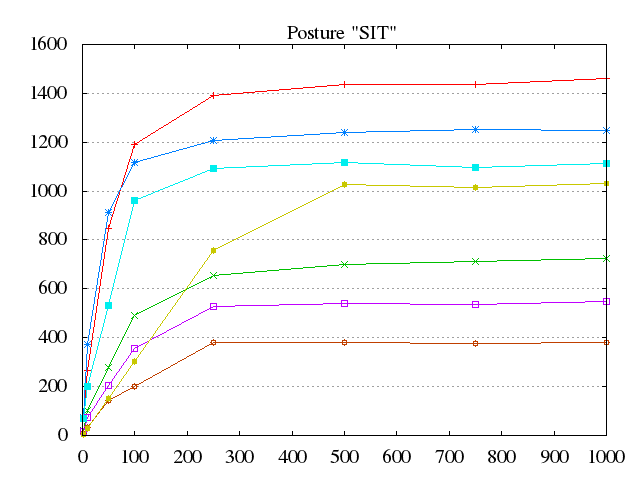}
 \end{subfigure}
 \begin{subfigure}{0.8\columnwidth}
 \centering
 \includegraphics[width=\textwidth,height=3cm]{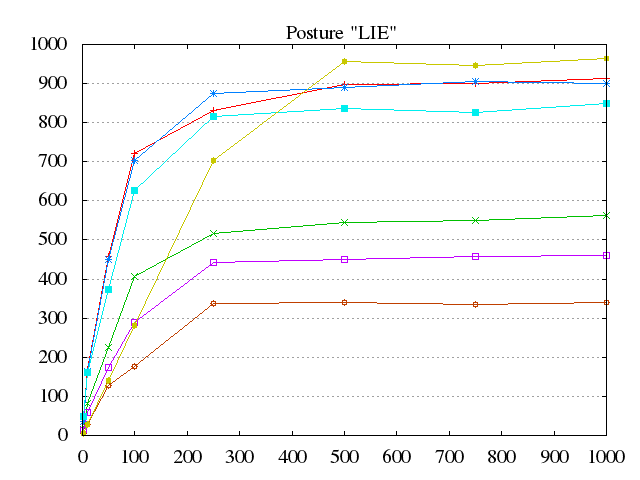}
 \end{subfigure}
 \begin{subfigure}{0.8\columnwidth}
 \centering
 \includegraphics[width=\textwidth,height=3cm]{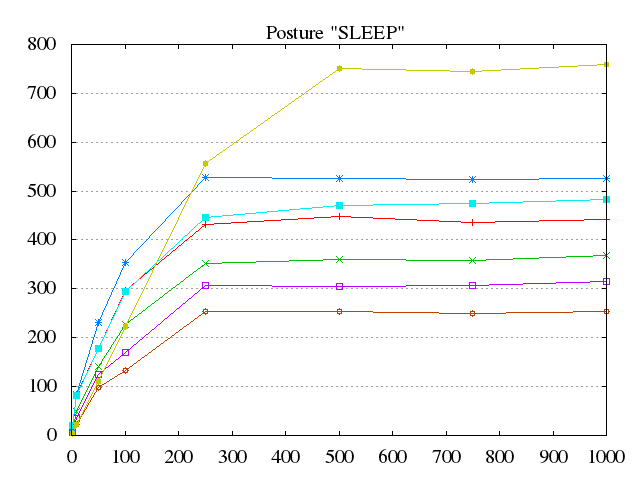}
 \end{subfigure}
 \begin{subfigure}{0.8\columnwidth}
 \centering
 \includegraphics[width=\textwidth,height=4.35cm]{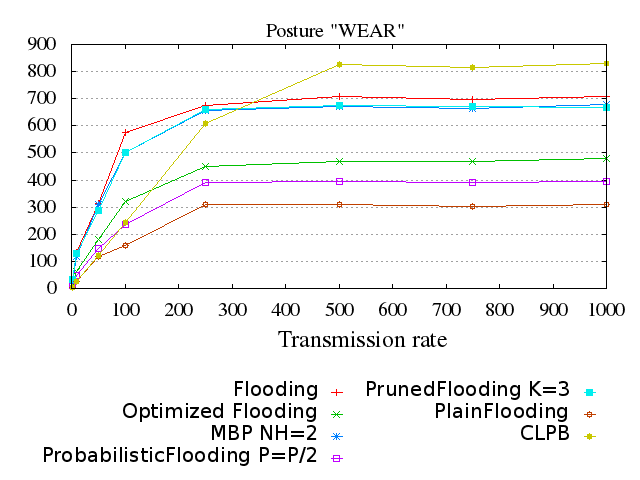}
 \end{subfigure}
 \caption{Total number of transmissions and receptions in function of transmission rate Buffer Size 100}
\label{TxRxStressAll100}
\end{figure}

Figure \ref{TxRxStressAll100} shows the average number of transmissions and receptions per node. 

For all strategies, the number of transmissions and receptions starts increasing with transmission rate. However, and contrary to expectations, from a given rate, transmissions and receptions number stagnates and remains so even if transmission rate increases.
%Nodes are unable to manage more packets. 
For broadcast strategies, this explains the low percentage of covered nodes: transmission rate increases while nodes capacity to handle packets is reached. 
Due to a high mobility and appearance of several links, with RUN posture, nodes exchange three times more packets than in SLEEP posture and this for all strategies.

\emph{CLPB} high transmissions and receptions number is due to a high percentage of covered nodes.

%%%%%%%%%%%%%%%%%%%%%%%%%%%%%%%%%%%%%%%%%%%%%%%
\subsection{Broadcast rates up to 1000 packets/s and buffer size 200}
%\label{BufferSize200}
MAC buffer size is set to $200$ packets. Now, more packets are able to be buffered waiting for broadcast. 

\subsubsection{Percentage of covered Nodes:}
\begin{figure}[htbp]
\centering
 \begin{subfigure}{0.8\columnwidth}
 \centering
 \includegraphics[width=\textwidth,height=3cm]{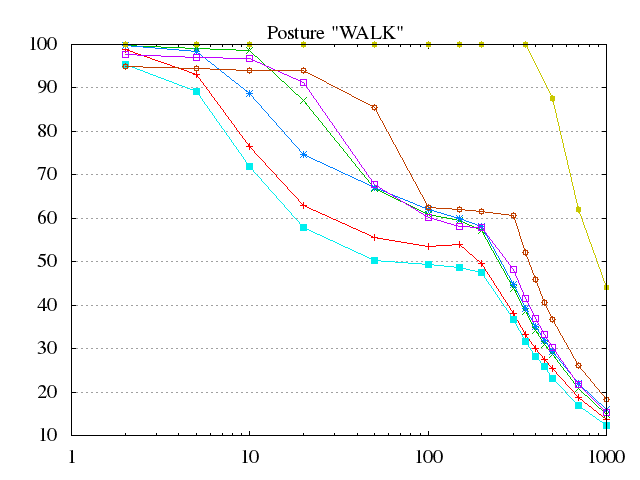}
 \end{subfigure}
 \begin{subfigure}{0.8\columnwidth}
 \centering
 \includegraphics[width=\textwidth,height=3cm]{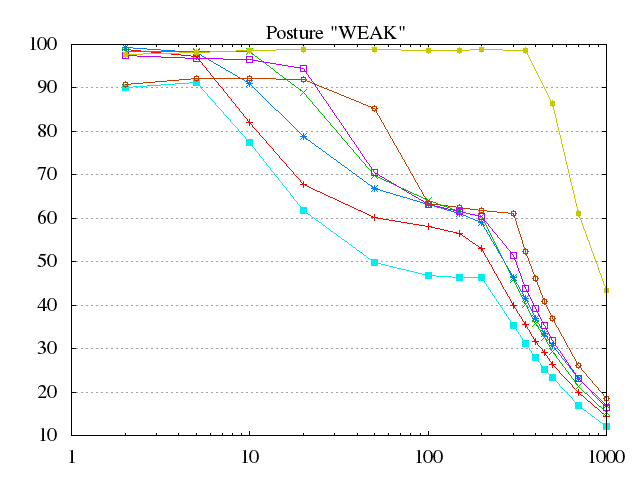}
 \end{subfigure}
 \begin{subfigure}{0.8\columnwidth}
 \centering
 \includegraphics[width=\textwidth,height=3cm]{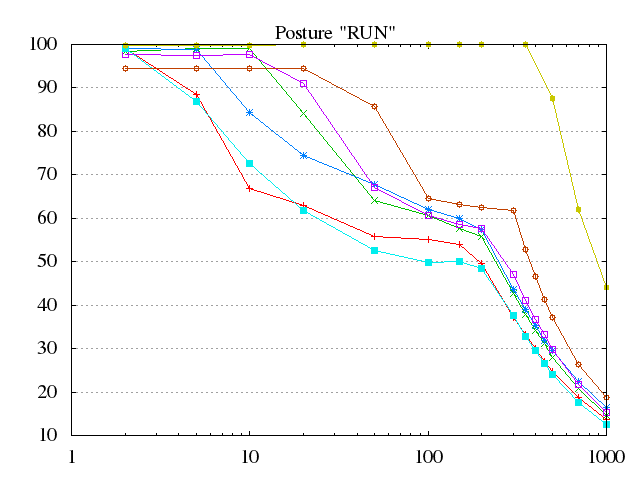}
 \end{subfigure}
 \begin{subfigure}{0.8\columnwidth}
 \centering
 \includegraphics[width=\textwidth,height=3cm]{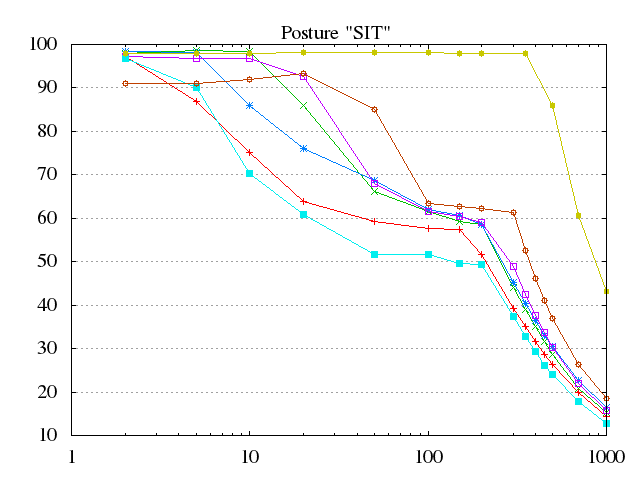}
 \end{subfigure}
 \begin{subfigure}{0.8\columnwidth}
 \centering
 \includegraphics[width=\textwidth,height=3cm]{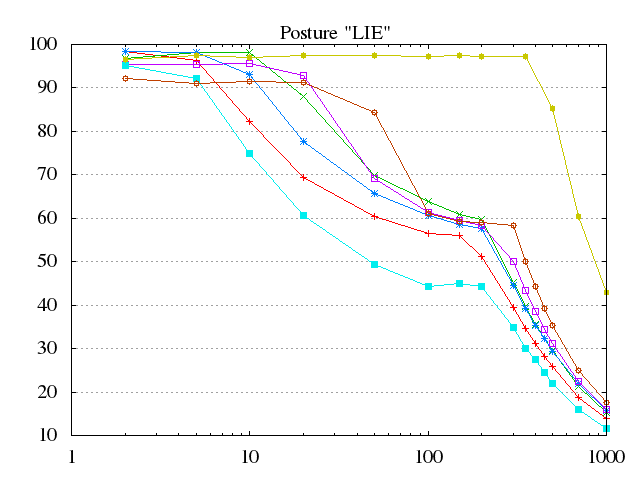}
 \end{subfigure}
 \begin{subfigure}{0.8\columnwidth}
 \centering
 \includegraphics[width=\textwidth,height=3cm]{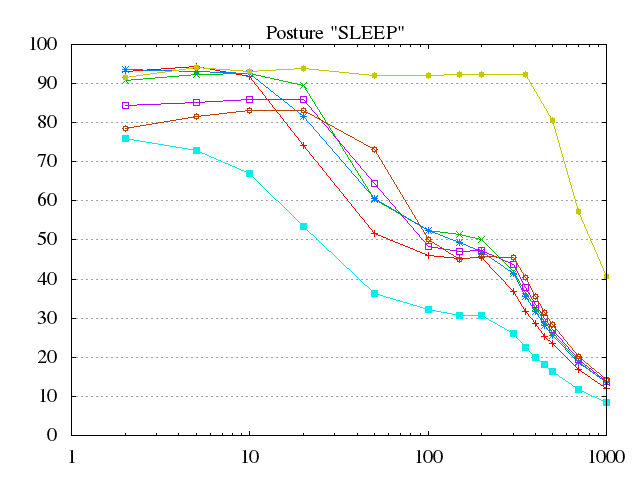}
 \end{subfigure}
 \begin{subfigure}{0.8\columnwidth}
 \centering
 \includegraphics[width=\textwidth,height=4.35cm]{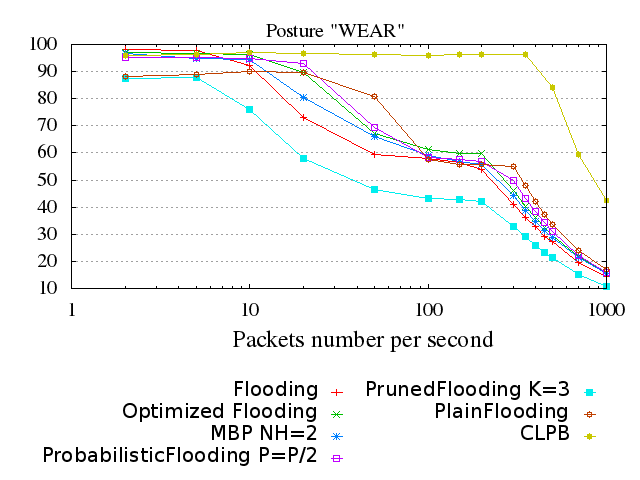}
 \end{subfigure}
 \caption{Percentage of covered nodes in function of transmission rate Buffer Size 200}
\label{PercentStressAll200}
\end{figure}

Figure \ref{PercentStressAll200} presents the percentage of covered nodes in function of transmissions rate.

Network coverage is no longer assured with increased transmission rate. \emph{Flooding} and \emph{Pruned Flooding} strategies show the lowest percentage due to collisions for both and to the random choice of next hops for \emph{Pruned Flooding}. \emph{Plain Flooding} slightly exceeds the other strategies. In this strategy, nodes broadcast received packets only once, thus number of packets copies in the network is limited. This points out again the impact of collisions. \emph{Optimized Flooding} and \emph{Probabilistic Flooding} performance is close to each others. These two strategies restrict rebroadcasting that allow them to have a bit better results compared to other strategies. However, we can observe, that, to $10$ packets/s, \emph{Optimized Flooding} slightly exceeds \emph{Probabilistic Flooding}. This gap is more important in SLEEP posture than RUN and WALK postures. Because, \emph{Optimized Flooding} strategy eliminates unnecessary retransmissions towards the end broadcasting process while \emph{Probabilistic Flooding} restricts broadcasting from the beginning.
\emph{CLPB} outperforms the other strategies with a high percentage of covered. By $1000$ packets/s, \emph{CLPB} covers $40\%$ of nodes against an average of $10\%$ for broadcast strategies.

\subsubsection{Percentage of De-sequencing:}
\begin{figure}[htbp]
\centering
 \begin{subfigure}{0.8\columnwidth}
 \centering
 \includegraphics[width=\textwidth,height=3cm]{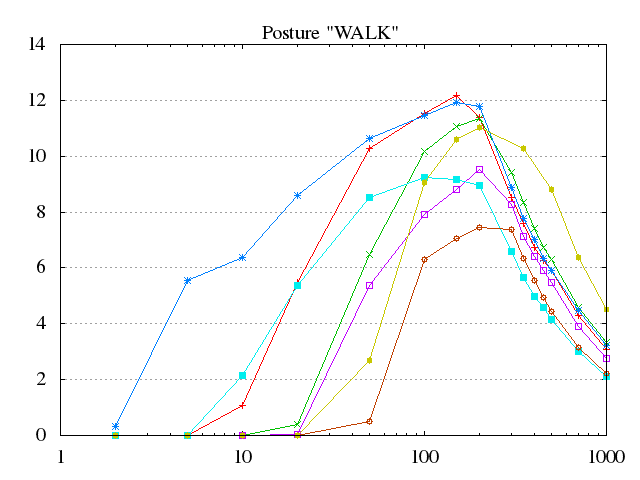}
 \end{subfigure}
 \begin{subfigure}{0.8\columnwidth}
 \centering
 \includegraphics[width=\textwidth,height=3cm]{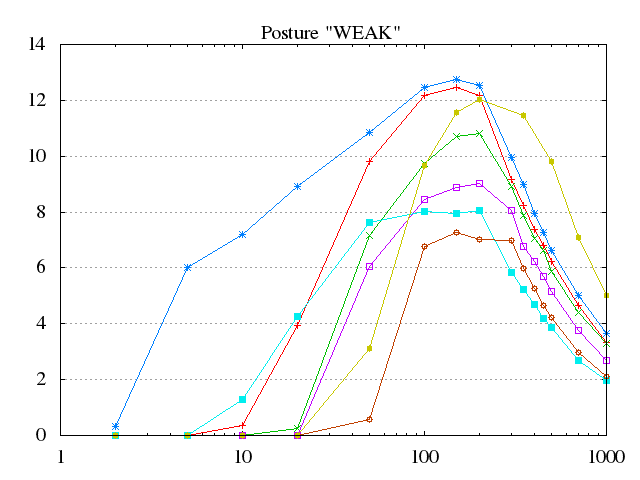}
 \end{subfigure}
 \begin{subfigure}{0.8\columnwidth}
 \centering
 \includegraphics[width=\textwidth,height=3cm]{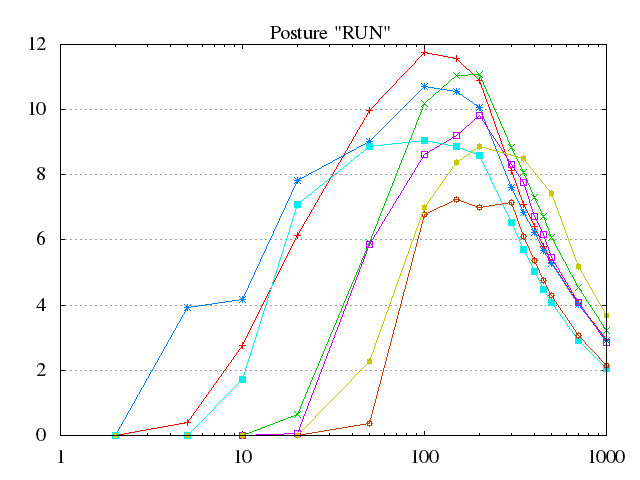}
 \end{subfigure}
 \begin{subfigure}{0.8\columnwidth}
 \centering
 \includegraphics[width=\textwidth,height=3cm]{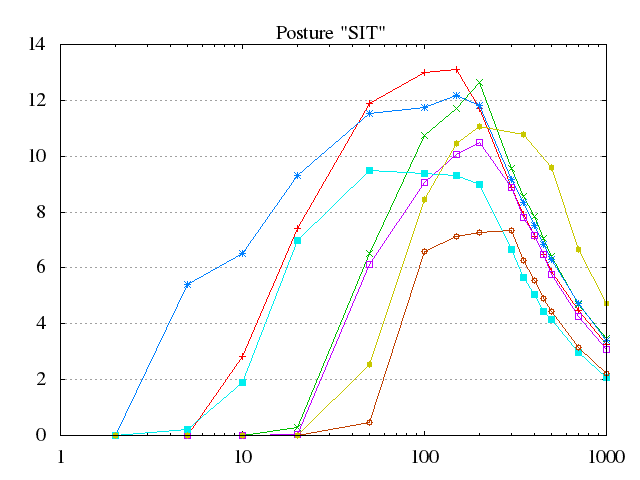}
 \end{subfigure}
 \begin{subfigure}{0.8\columnwidth}
 \centering
 \includegraphics[width=\textwidth,height=3cm]{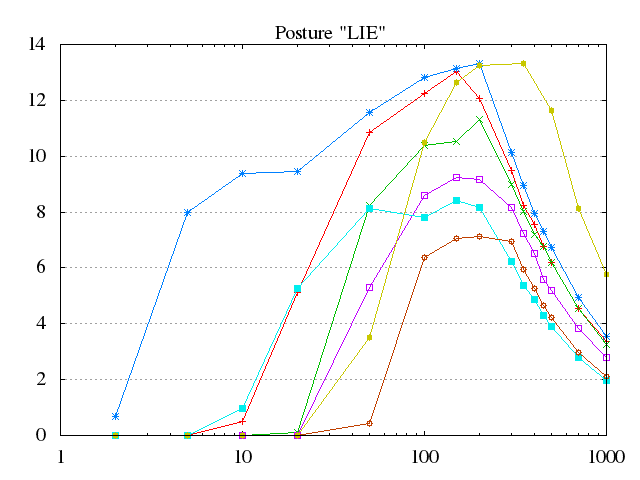}
 \end{subfigure}
 \begin{subfigure}{0.8\columnwidth}
 \centering
 \includegraphics[width=\textwidth,height=3cm]{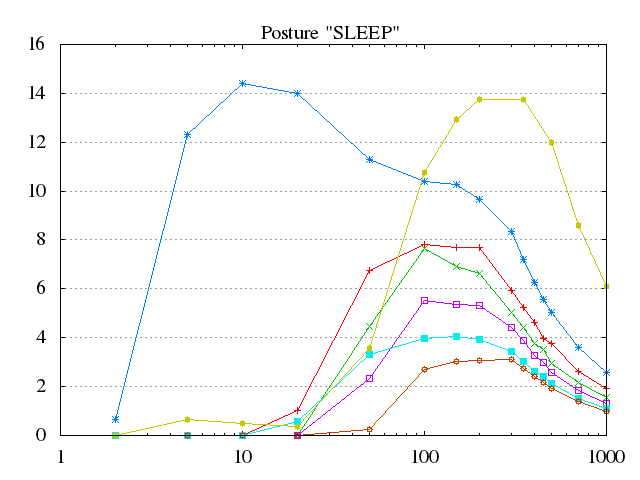}
 \end{subfigure}
 \begin{subfigure}{0.8\columnwidth}
 \centering
 \includegraphics[width=\textwidth,height=4.35cm]{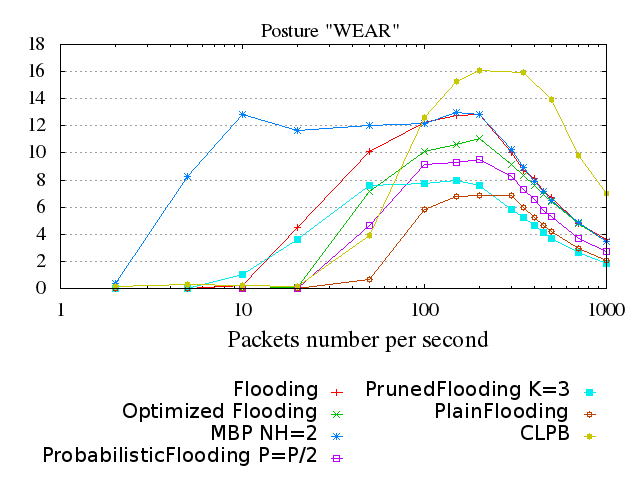}
 \end{subfigure}
 \caption{Percentage of De-sequencing in function of transmission rate Buffer Size 200}
\label{DeseqStressAll200}
\end{figure}

Figure \ref{DeseqStressAll200} presents the percentage of de-sequencing in function of packets number per second. 

At first glance, results are quite similar to results for a buffer size equal to $100$ (section \ref{Deseq100}). Indeed, the percentage of de-sequencing is close for both values. However, in details, results show that the peak reached in percentage of de-sequencing is offset and around $200$ packets/s whereas it is around $100$ packets/s for the previous results.
With \emph{CLPB}, the peak reached is reached by $350$ packets/s, starting from which percentage of both covered nodes and de-sequencing decreases.

\subsubsection{Number of transmissions and receptions}
\begin{figure}[htbp]
\centering
 \begin{subfigure}{0.8\columnwidth}
 \centering
 \includegraphics[width=\textwidth,height=3.2cm]{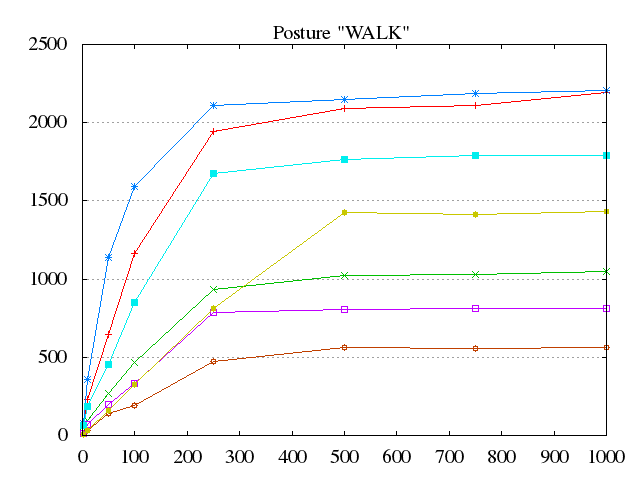}
 \end{subfigure}
 \begin{subfigure}{0.8\columnwidth}
 \centering
 \includegraphics[width=\textwidth,height=4.85cm]{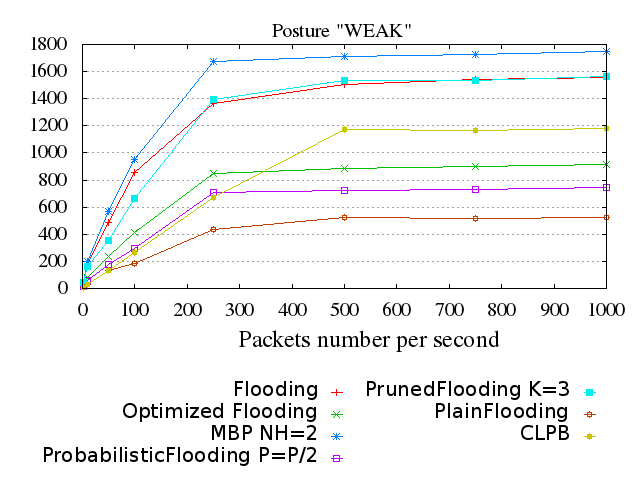}
 \end{subfigure}
 \caption{Total number of transmissions and receptions in function of transmission rate Buffer Size 200}
\label{TxRxStressAll200}
\end{figure}

Figure \ref{TxRxStressAll200} shows the average number of transmissions and receptions For WLAK and WEAK postures. 

The number increases with transmission rate increase and stagnates by $200$ packets/s. This value is equal to MAC buffer size. It shows again the impact of buffer size on strategies performance. This stagnation shows that the nodes are unable to manage more packets in the network. This explains too percentage of covered nodes decrease: transmission rate increases while nodes capacity to handle packets is reached.

%%%%%%%%%%%%%%%%%%%%%%%%%%%%%%%%%%%%%%%%%%%%%%%
\subsection{Broadcast rates up to 1000 packets/s and buffer size 300}
\label{BufferSize300}
%MAC buffer size is set to $300$ packets. Now, nodes are able to put in buffer more packets waiting to be broadcasted. As for buffer size $100$ (paragraph \ref{BufferSize100}), we present the percentage of covered nodes, the percentage of de-sequencing and the average number of transmissions and receptions per node.
In this section we advocate that, contrary to the expectations, the increase of the buffer size (three times greater than in the previous case) has little impact on the performances of both flat and cross-layer broadcast.

\subsubsection{Percentage of covered Nodes}
\begin{figure}[htbp]
\centering
 \begin{subfigure}{0.8\columnwidth}
 \centering
 \includegraphics[width=\textwidth,height=3cm]{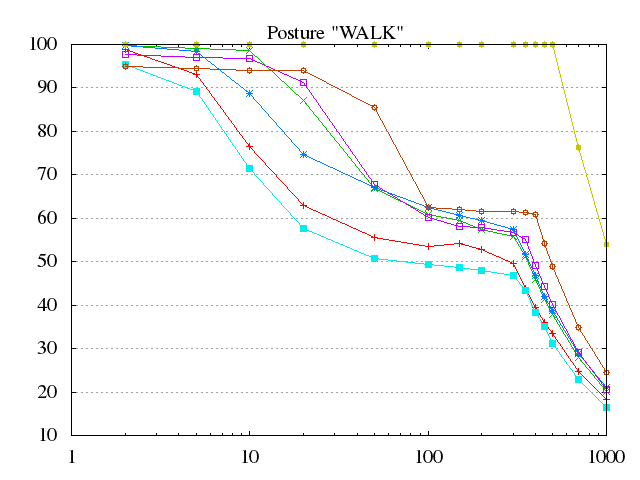}
 \end{subfigure}
 \begin{subfigure}{0.8\columnwidth}
 \centering
 \includegraphics[width=\textwidth,height=3cm]{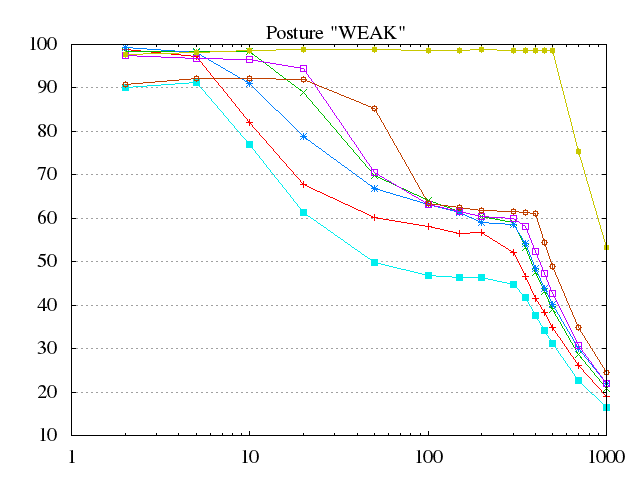}
 \end{subfigure}
 \begin{subfigure}{0.8\columnwidth}
 \centering
 \includegraphics[width=\textwidth,height=3cm]{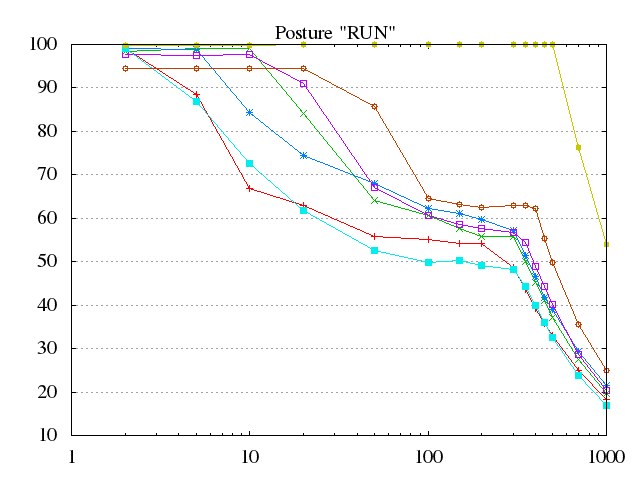}
 \end{subfigure}
 \begin{subfigure}{0.8\columnwidth}
 \centering
 \includegraphics[width=\textwidth,height=3cm]{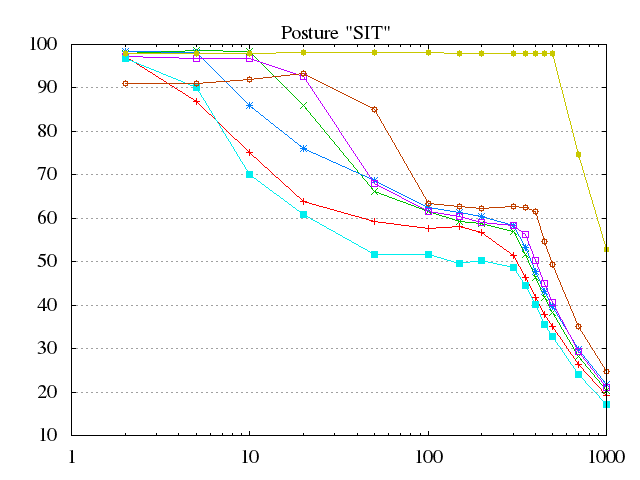}
 \end{subfigure}
 \begin{subfigure}{0.8\columnwidth}
 \centering
 \includegraphics[width=\textwidth,height=3cm]{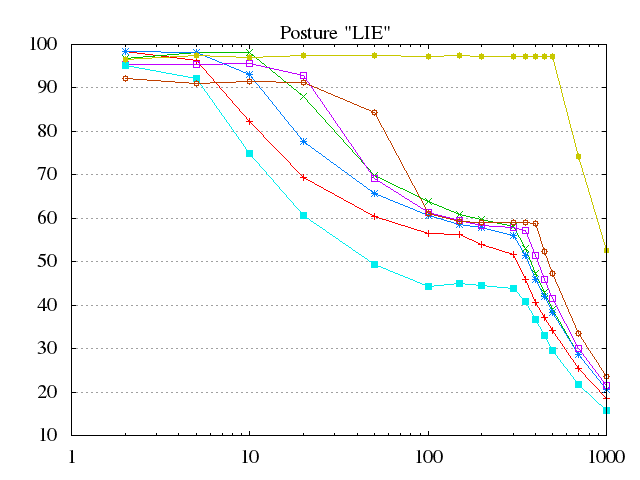}
 \end{subfigure}
 \begin{subfigure}{0.8\columnwidth}
 \centering
 \includegraphics[width=\textwidth,height=3cm]{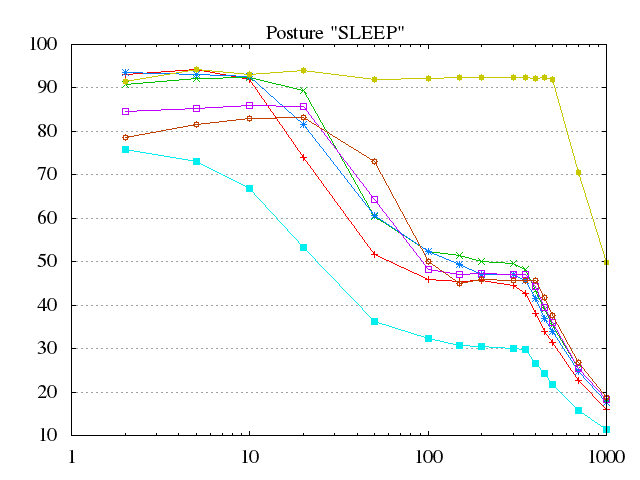}
 \end{subfigure}
 \begin{subfigure}{0.8\columnwidth}
 \centering
 \includegraphics[width=\textwidth,height=4.35cm]{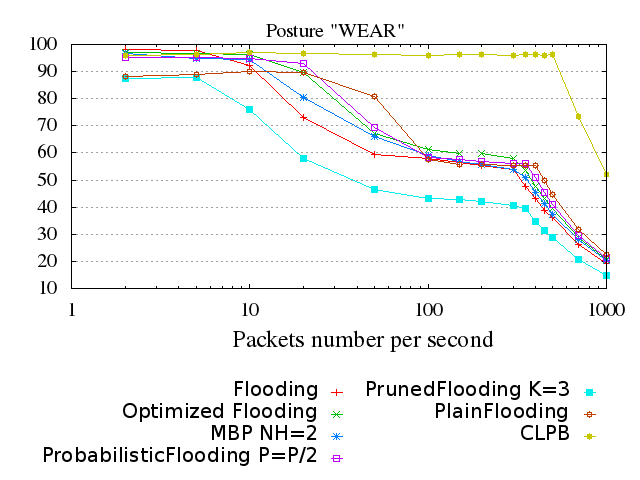}
 \end{subfigure}
 \caption{Percentage of covered nodes in function of transmission rate Buffer Size 300}
\label{PercentStressAll300}
\end{figure}

Figure \ref{PercentStressAll300} presents the percentage of covered nodes in function of transmission rate.

\emph{Flooding} and \emph{Pruned Flooding} strategies have the lowest percentage due to collisions. For \emph{Pruned Flooding}, it is also due to the random choice of next hops. \emph{Plain Flooding} slightly exceeds the other strategies in WALK and RUN postures. In this strategies, nodes broadcast received packets only once, thus number of packets copies in the network is limited.
 This behavior is once again due to collisions.
 However in SLEEP posture (which is a static posture) \emph{Plain Flooding} maximum percentage is $83\%$ against $94\%$ in RUN and WEAK postures. \emph{Optimized Flooding} and \emph{Probabilistic Flooding} performances are close to each others. These two strategies restrict rebroadcasting which avoid unnecessary retransmissions and gives a better percentage compared to other strategies. However, in SLEEP posture, up to $10$ packets/s, \emph{Optimized Flooding} exceeds \emph{Probabilistic Flooding}. This is due to the fact that \emph{Optimized Flooding} strategy eliminates unnecessary retransmissions towards the end of broadcasting process while \emph{Probabilistic Flooding} restricts broadcasting from the beginning.

\emph{\emph{CLPB} outperforms all broadcast strategies.It shows high percentage of covered nodes equal to $100\%$ for WALK and RUN postures and equal to $92\%$ for SLEEP posture up to $500$ packets/s. When the transmission rate is $1000$ packets/s, \emph{CLPB} covers $50\%$ of nodes (half the network) against an average of $20\%$ for flat broadcast strategies. }

\subsubsection{Percentage of De-sequencing}
\begin{figure}[htbp]
\centering
 \begin{subfigure}{0.8\columnwidth}
 \centering
 \includegraphics[width=\textwidth,height=3cm]{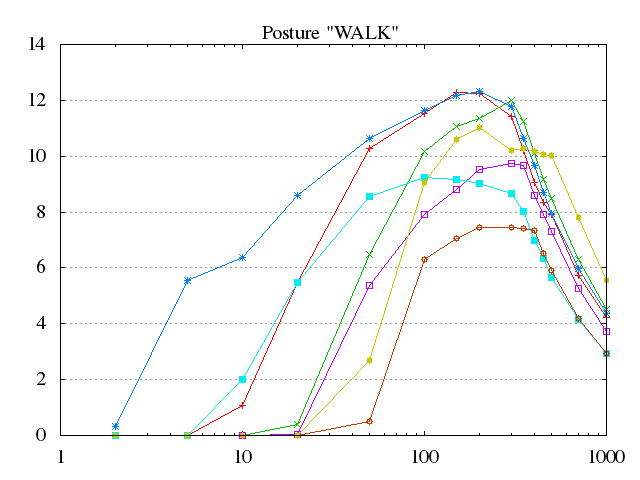}
 \end{subfigure}
 \begin{subfigure}{0.8\columnwidth}
 \centering
 \includegraphics[width=\textwidth,height=3cm]{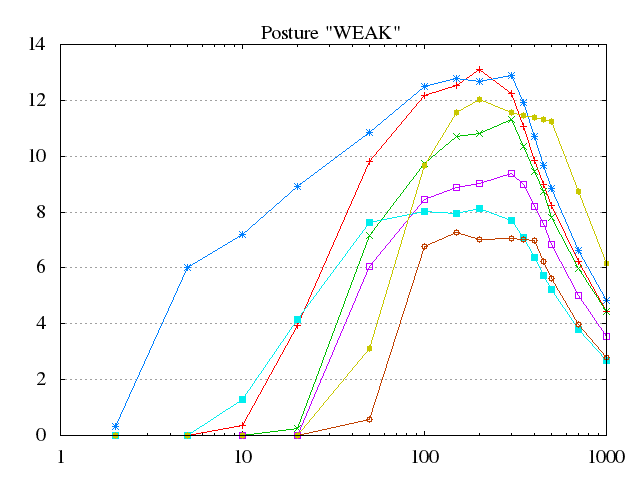}
 \end{subfigure}
 \begin{subfigure}{0.8\columnwidth}
 \centering
 \includegraphics[width=\textwidth,height=3cm]{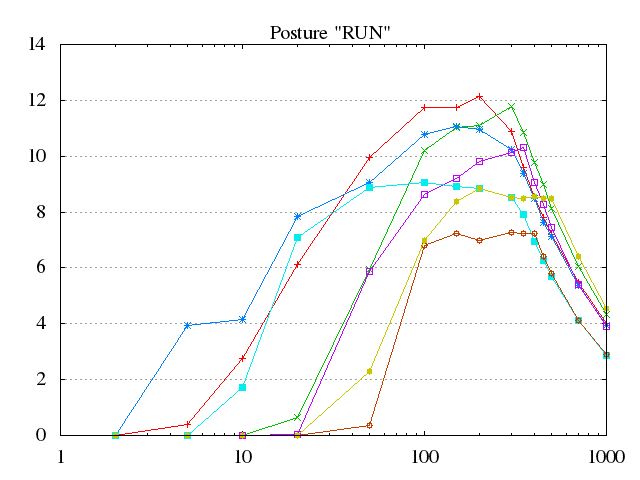}
 \end{subfigure}
 \begin{subfigure}{0.8\columnwidth}
 \centering
 \includegraphics[width=\textwidth,height=3cm]{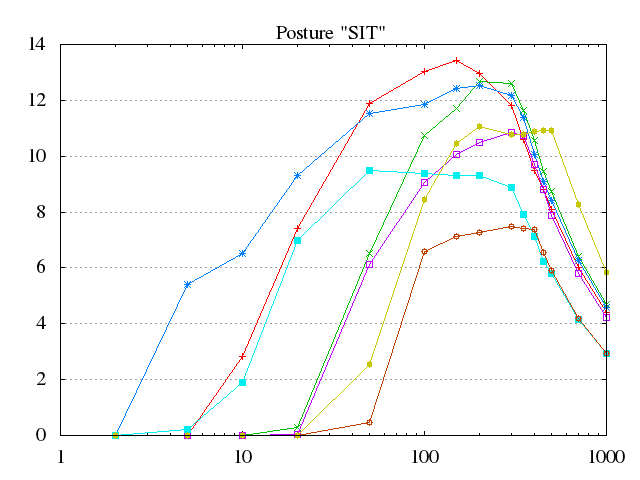}
 \end{subfigure}
 \begin{subfigure}{0.8\columnwidth}
 \centering
 \includegraphics[width=\textwidth,height=3cm]{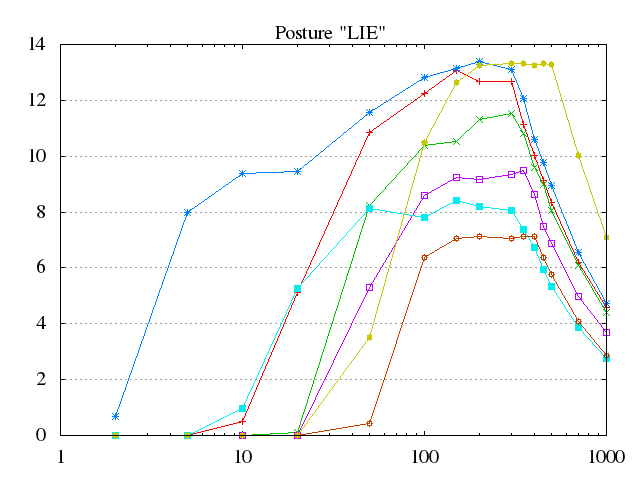}
 \end{subfigure}
 \begin{subfigure}{0.8\columnwidth}
 \centering
 \includegraphics[width=\textwidth,height=3cm]{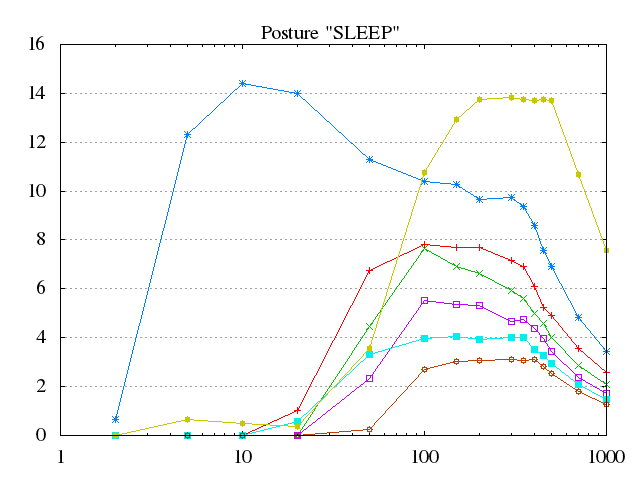}
 \end{subfigure}
 \begin{subfigure}{0.8\columnwidth}
 \centering
 \includegraphics[width=\textwidth,height=4.35cm]{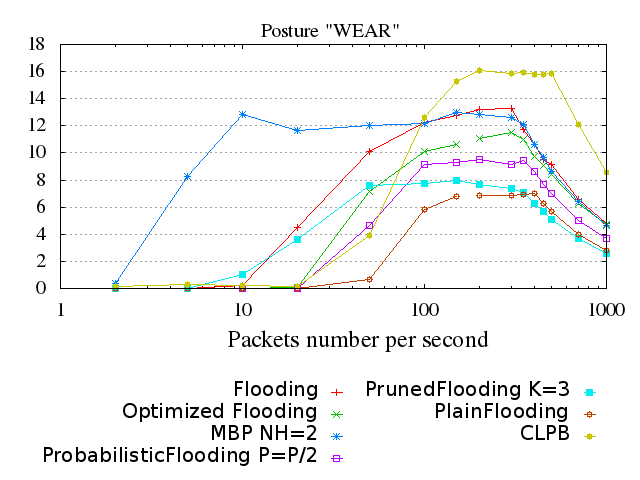} 
 \end{subfigure}
 \caption{Percentage of De-sequencing in function of transmission rate Buffer Size 300}
\label{DeseqStressAll300}
\end{figure}

Figure \ref{DeseqStressAll300} presents the percentage of de-sequencing function of transmission rate. 

At first glance, results are quite similar to results for buffer size $100$ (paragraph \ref{Deseq100}). Indeed, the percentage of de-sequencing is close for both buffer size values in a range of [$3-14$]\%. However, looking in details, results show that the curve is offset and the percentage of de-sequencing starts decreasing around $300$ packets/s whereas it is around $100$ packets/s for the previous results. Looking back to the percentage of covered nodes, it falls linearly by $300$ packets/s, at this point, packets from application layer are not even broadcasted due to MAC over-buffer.

\subsubsection{Average number of transmissions and receptions per node}
\begin{figure}[htbp]
\centering
 \begin{subfigure}{0.8\columnwidth}
 \includegraphics[width=\textwidth,height=3cm]{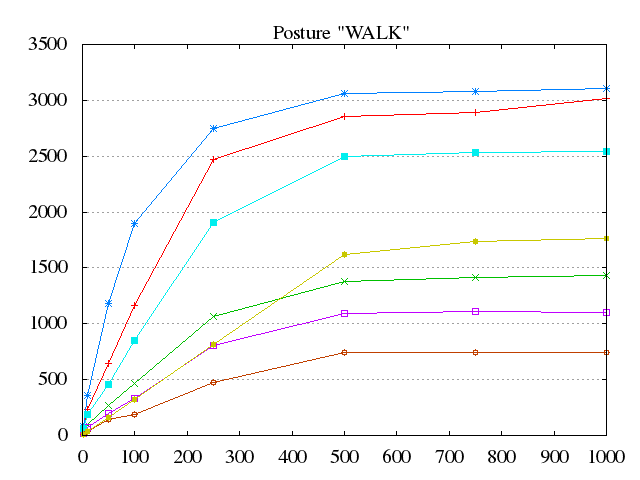}
 \end{subfigure}
 \begin{subfigure}{0.8\columnwidth}
 \includegraphics[width=\textwidth,height=3cm]{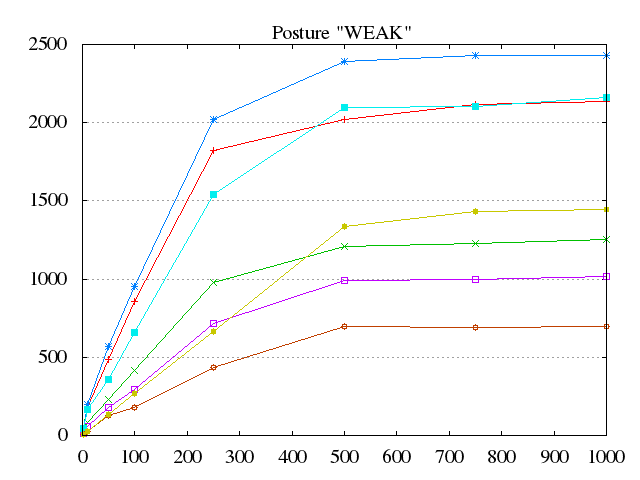}
 \end{subfigure}
 \begin{subfigure}{0.8\columnwidth}
 \centering
 \includegraphics[width=\textwidth,height=3cm]{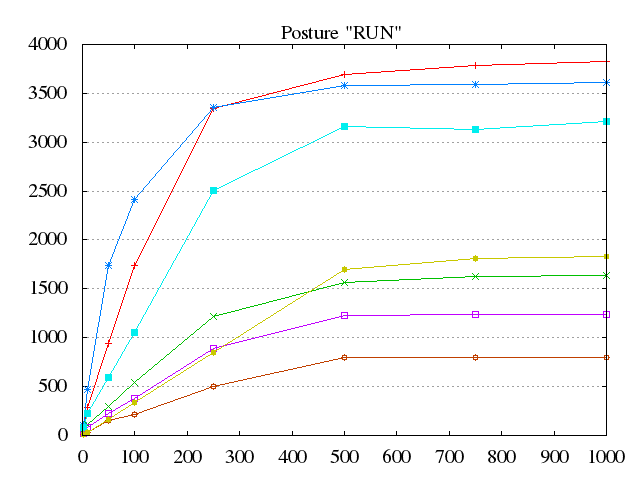}
 \end{subfigure}
 \begin{subfigure}{0.8\columnwidth}
 \centering
 \includegraphics[width=\textwidth,height=3cm]{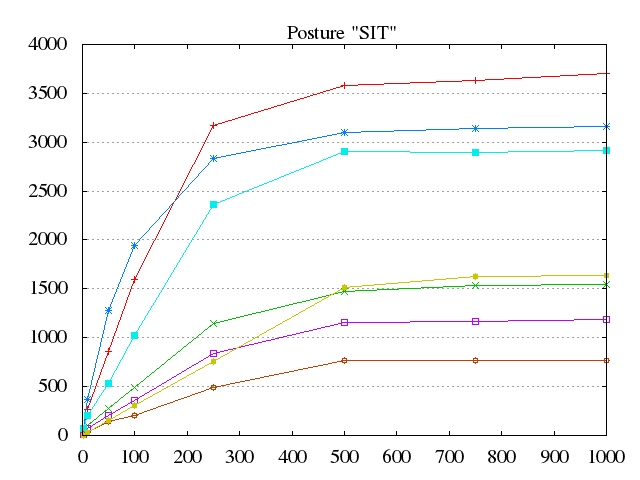}
 \end{subfigure}
 \begin{subfigure}{0.8\columnwidth}
 \centering
 \includegraphics[width=\textwidth,height=3cm]{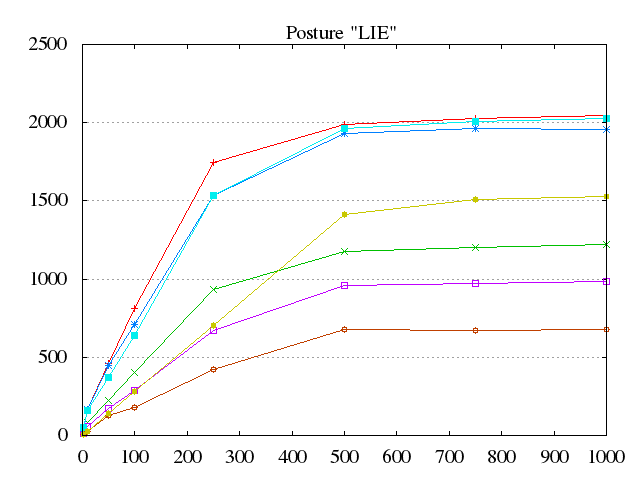}
 \end{subfigure}
 \begin{subfigure}{0.8\columnwidth}
 \centering
 \includegraphics[width=\textwidth,height=3cm]{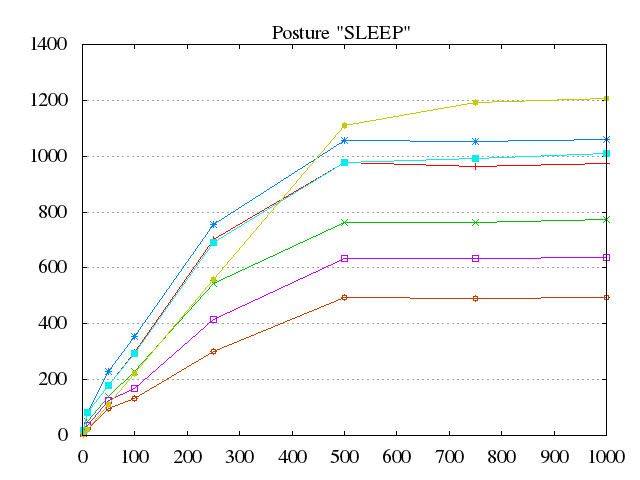}
 \end{subfigure}
 \begin{subfigure}{0.8\columnwidth}
 \centering
 \includegraphics[width=\textwidth,height=4.35cm]{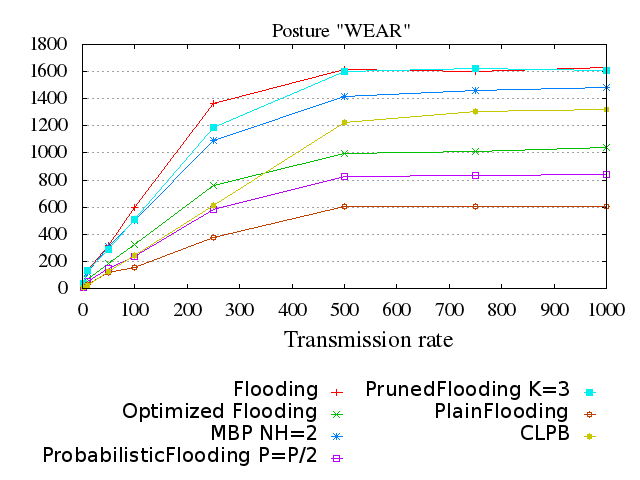}
 \end{subfigure}
 \caption{Average number of transmissions and receptions per node in function of transmission rate Buffer Size 300}
\label{TxRxStressAll300}
\end{figure}

Figure \ref{TxRxStressAll300} shows the average number of transmissions and receptions per node.

The number increases and stabilizes around $300$ packets/s. It shows again the impact of buffer size on strategies performance.

\emph{Interesting results for postures RUN and WALK points out that, up to $500$ packets/s, \emph{CLPB} transmissions and receptions number is equal to half \emph{MBP} and \emph{Flooding} transmissions and receptions number. However, \emph{CLPB} ensures $100\%$ of covered nodes while these two strategies only cover $35\%$ of the network.}

These are important results that, contrary to expectations, show that, although the buffer size is three times greater, broadcast strategies performance are barely better than performance with buffer size set to $100$ (see Section \ref{CLPBperf}). Percentage of covered nodes decreases to $20\%$ by $1000$ instead of $10\%$. As a result of this slight increase, percentage of de-sequencing increases. Also, the percentage of covered nodes falls when transmission rates exceed $10$ packets/s in average and no improvement is observed. 

\emph{\emph{CLPB} outperforms all broadcast strategies and maintains a higher percentage of covered nodes than with a buffer size set to $100$ up to $500$ packets/s.}

%%%%%%%%%%%%%%%%%%%%%%%%%%%%%%%%%%%%%%%%%%%%%%%
%\newpage
\subsection{Broadcast with transmission rates of 1 packet/s} 
\label{NewVsOld}
The previous study (sections \ref{CLPBperf} and \ref{BufferSize300}) showed that our new cross layer protocol, \emph{CLPB}, outperforms flat broadcast strategies with a \emph{Sink} transmission rate goes up to 1000 packets/s. The protocol offers a high percentage of covered nodes up to $500$ packets/s against $20$ packets/s for broadcast strategies (For buffer size $300$).

In the following, we zoom the case when \emph{Sink} transmission rate equals one packet per second. In addition to the percentage of covered nodes and the number of transmissions and receptions we also evaluate the \emph{end-to-end delay} per strategy.

\subsubsection{Percentage of covered nodes}

Figure \ref{PercentCrossAll} shows the average percentage of covered nodes for all strategies.

\begin{figure}[htbp]
\centering
 \begin{subfigure}{0.8\columnwidth}
 \centering
 \includegraphics[width=\textwidth,height=4.5cm]{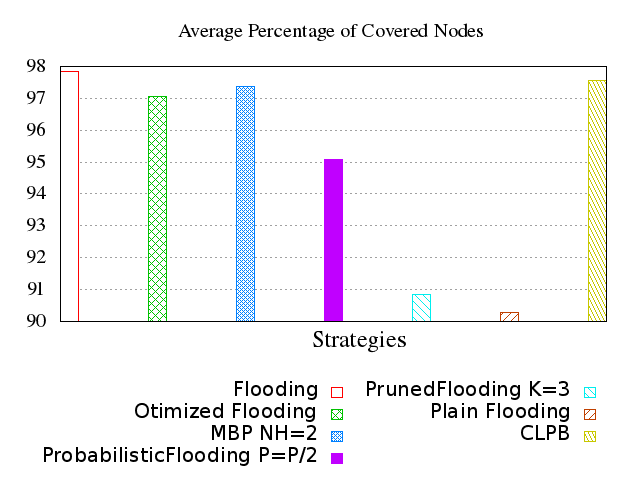}
 \end{subfigure}
 \caption{Average Percentage of Covered Nodes}
\label{PercentCrossAll}
\end{figure}

Results show, in average, a good percentage for all strategies greater than $90\%$.

\emph{Flooding}, \emph{Optimized Flooding} and \emph{MBP} show quite close percentages. As expected, with \emph{Flooding} strategy, nodes rebroadcast each received copy of the packet which increases the probability of receiving the packet.
\emph{Optimized Flooding} is a \emph{Flooding} strategy adaptation to meet \emph{Flooding} strategy advantages with WBAN energy constraints. Thus, a slight decrease of the percentage (less than $2\%$) is observed due to discarding obsolete copies of the packet in order to reduce ping-pong effect.
For \emph{MBP}, nodes broadcast each received packet while the number of hops of this packet is less than the predefined threshold (in these simulations: threshold=NH=$2$). If threshold is reached, nodes delay packet broadcasting for a moment. Thus, a packet is only discarded when a sender receives required acknowledgements otherwise it continues broadcasting waiting packets. This explains the good percentage shown by \emph{MBP}.

\emph{Probabilistic Flooding} and \emph{Pruned Flooding} strategies use randomness in the broadcast process: \emph{Probabilistic Flooding} in the choice of random variable $r$ to compare with the probability $P$ while \emph{Pruned Flooding} in the choice of random nodes as next hops. \emph{Pruned Flooding} is more affected by body postures than \emph{Probabilistic Flooding}. It shows the lowest percentage of covered nodes in SLEEP posture equal to $74.6\%$ while \emph{Probabilistic Flooding} percentage is equal to $90.3\%$. This reveals the importance of forwarding nodes position on body to the number of packet copies in the network.

\emph{Plain Flooding} shows the lowest percentage. Nodes broadcast a received packet only once, the first time it receives it. After that, all received copies are discarded. When a neighbor didn't receive the packet at that time, it has less chances to receive it after.

\emph{\emph{CLPB} shows a high percentage, higher than $97\%$. Although nodes broadcast the packet only once, \emph{CLPB} achieves good network cover.}

\begin{figure}[htbp]
\centering
 \begin{subfigure}{0.9\columnwidth}
 \centering
 \includegraphics[width=\textwidth,height=3.05cm]{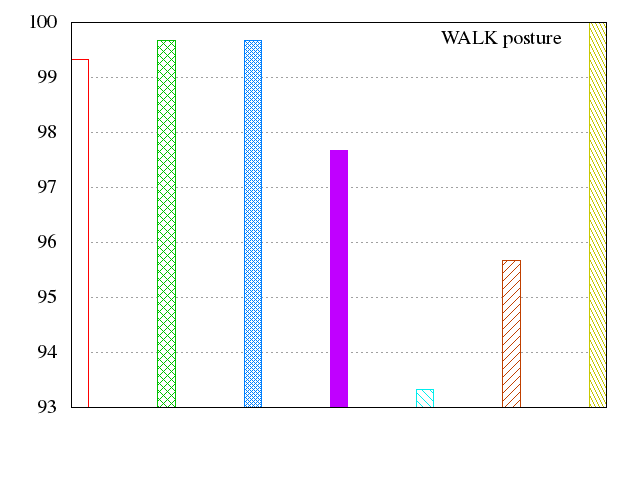}
 \end{subfigure}
 \begin{subfigure}{0.9\columnwidth}
 \centering
 \includegraphics[width=\textwidth,height=3.05cm]{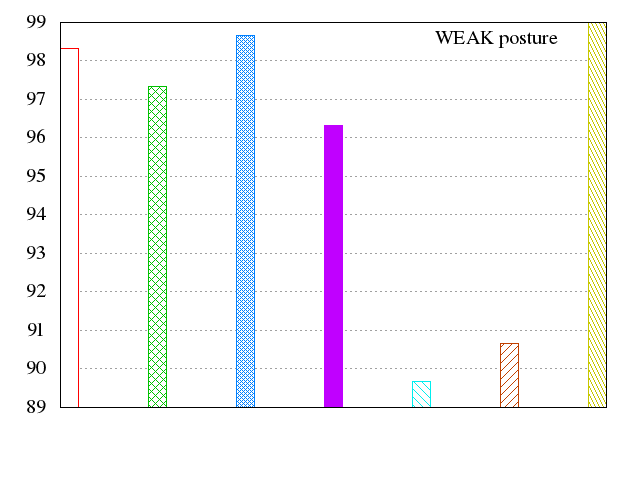}
 \end{subfigure}
 \begin{subfigure}{0.9\columnwidth}
 \centering
 \includegraphics[width=\textwidth,height=3.05cm]{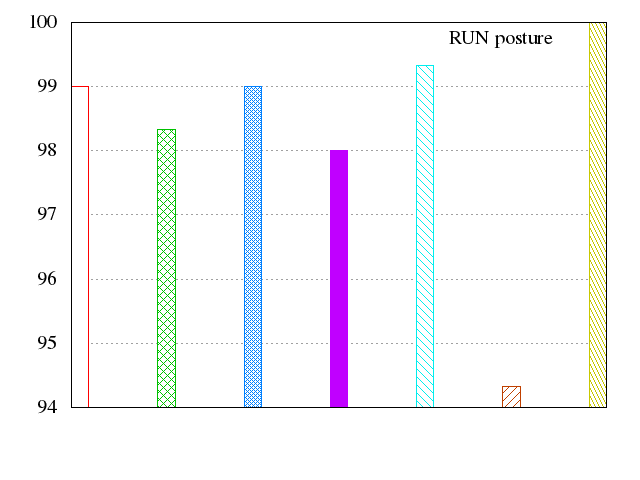}
 \end{subfigure}
 \begin{subfigure}{0.9\columnwidth}
 \centering
 \includegraphics[width=\textwidth,height=3.05cm]{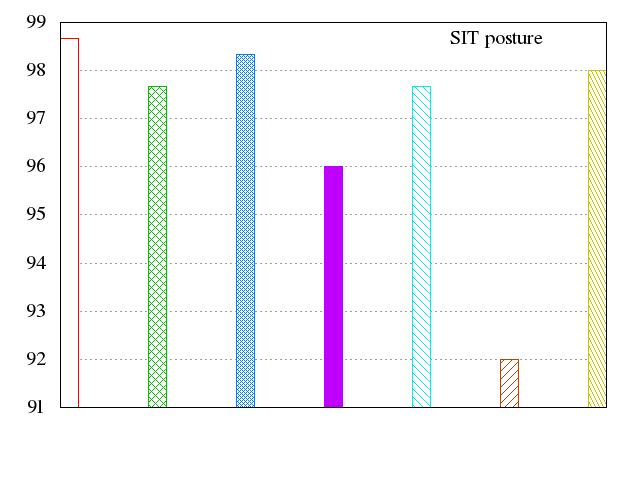}
 \end{subfigure}
 \begin{subfigure}{0.9\columnwidth}
 \centering
 \includegraphics[width=\textwidth,height=3.05cm]{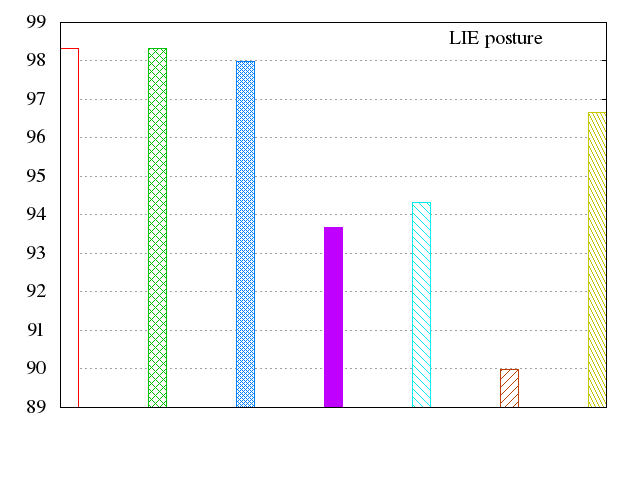}
 \end{subfigure}
 \begin{subfigure}{0.9\columnwidth}
 \centering
 \includegraphics[width=\textwidth,height=3.05cm]{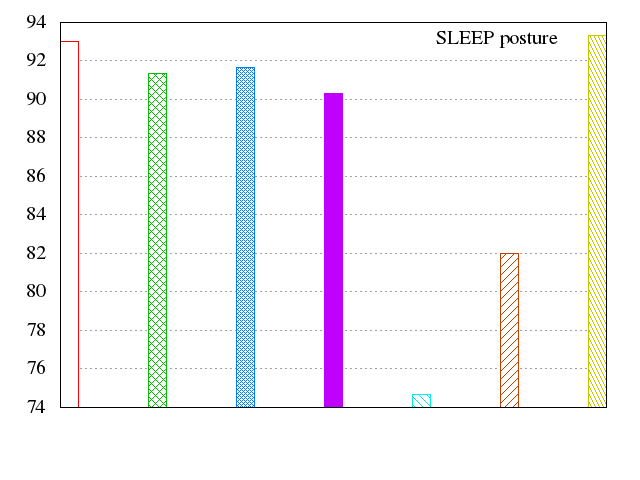}
 \end{subfigure}
 \begin{subfigure}{0.9\columnwidth}
 \centering
 \includegraphics[width=\textwidth,height=3.75cm]{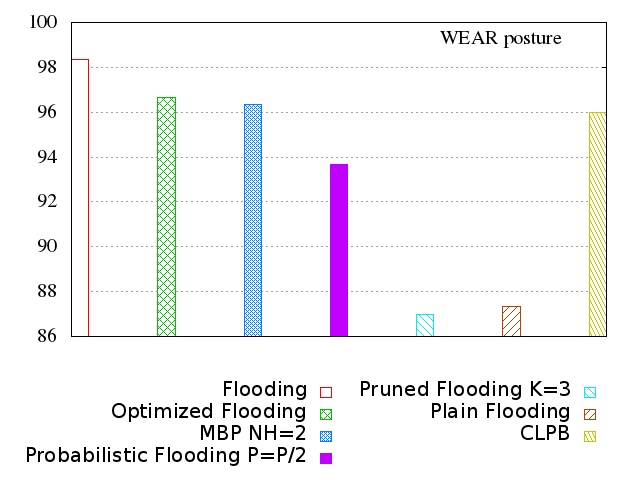}
 \end{subfigure}
 \caption{Percentage of covered nodes for all postures }
\label{PercentCrossAll}
\end{figure}

\paragraph{End to end delay}

\begin{figure}[htbp]
 \begin{subfigure}{0.8\columnwidth}
 \centering
 \includegraphics[width=\textwidth,height=4.5cm]{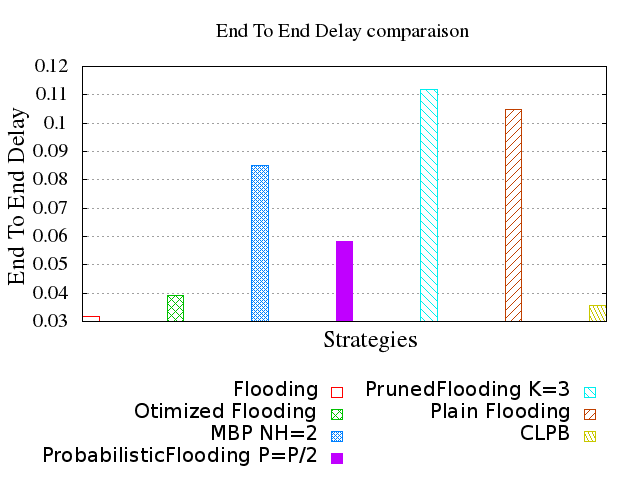}
 \end{subfigure}
 \caption{End To End Delay comparaison}
 \label{EEDAll}
\end{figure}

Figure \ref{EEDAll} shows the average end to end delay for all strategies. 
%Contrary to the percentage of covered nodes (Figure \ref{PercentCrossAll}), strategies performance varies remarkably.

\emph{Flooding} strategy has the lowest end to end delay and remains a reference regarding the delay to cover the hole network. Because no restrictions are imposed to broadcast, nodes automatically and immediately broadcast each received copy of the packet.

\emph{Optimized Flooding} also ensures a good end to end delay. Although, nodes discards some received packets, the average end to end delay is quite close to \emph{Flooding} delay.

In \emph{MBP} nodes delay packet broadcasting in order to allow receivers to acknowledge received packets before overloading the network with unnecessary packet copies.

The highest end-to-end delay is observed with \emph{Plain Flooding} and \emph{Pruned Flooding} strategies. It is related to their low percentage of covered nodes. In fact, to each non covered node, a high end to end delay is automatically assigned to this node. 

\emph{CLPB} ensures the lowest end to end delay just after \emph{Flooding}. Although, sender nodes have to wait for their time slot to broadcast a packet, delay is not affected. This is due to two factors: First, the choice of time slot duration, equal to $0.005$s during these simulations. Second, \emph{CLPB} scheduling mechanism avoids collisions, increases transmission success probability, decreases percentage on non-covered nodes and thus reduces end to end delay.

\begin{figure}[htbp]
 \begin{subfigure}{0.9\columnwidth}
 \centering
 \includegraphics[width=\textwidth,height=4.5cm]{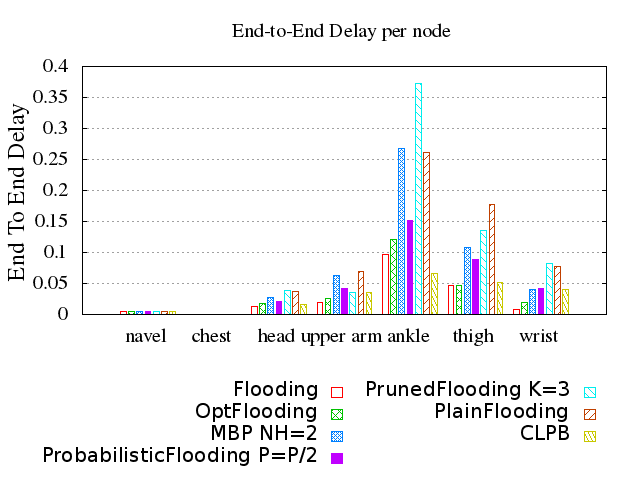}
 \end{subfigure}
 \caption{End To End Delay per node}
 \label{EEDNodesAll}
\end{figure}

Figure \ref{EEDNodesAll} shows the average end to end delay per node for all strategies. 

Strategies behave similarly considering nodes position. For nodes at navel and head, a low end to end delay is observed and it is almost equal for all strategies. In fact, these nodes are by a single hop from \emph{Sink} node adding to that, they are almost motionless. A slit increase in delay with upper arm and wrist nodes because, for some postures, these two nodes are moving and signal attenuation become important. Thigh and wrist, are the farthest nodes to the \emph{Sink}, thus packet needs to pass through intermediate nodes before reaching these two nodes which increases end to end delay. It is important to points out that, to reach ankle node, with our cross layer protocol a broadcasted packet takes half the time it takes with the other strategies even comparing with \emph{Flooding} strategy.

\begin{figure}[htbp]
 \begin{subfigure}{0.8\columnwidth}
 \centering
 \includegraphics[width=\textwidth,height=4.5cm]{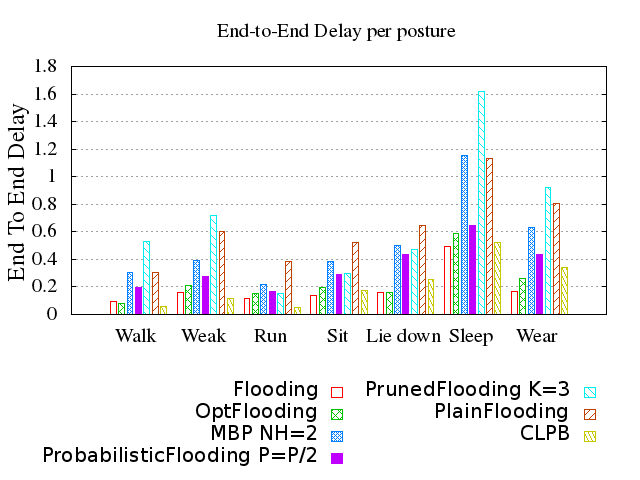}
 \end{subfigure}
 \caption{End To End Delay per posture}
 \label{EEDPosturesAll}
\end{figure}

Figure \ref{EEDPosturesAll} shows the average end to end delay per posture.

Our novel protocol \emph{CLPB} outperforms \emph{Flooding} in three postures: WALK, WEAK and RUN. These postures are variations of walking motions where links are reliable and signals attenuations are moderate. Even if it is considered as a reliable environment for communication, for \emph{Flooding}, it causes more packets exchange, collisions, packets loss and retransmissions.

With SLEEP posture, all strategies require more time to cover the network. In this posture, some nodes are hidden by body parts and, due to a motionless posture, they are difficult to reach.

\begin{figure}[htbp]
 \begin{subfigure}{0.8\columnwidth}
 \centering
 \includegraphics[width=\textwidth,height=4.5cm]{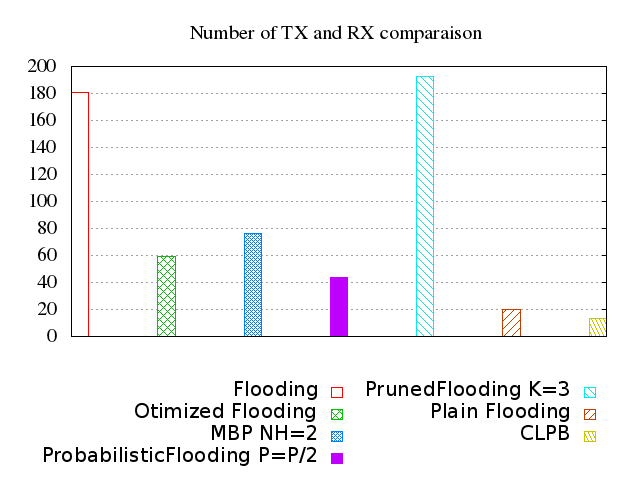}
 \end{subfigure}
 \caption{Transmissions and receptions number for all strategies}
 \label{TxRxAll}
\end{figure}

\paragraph{Total number of transmissions and receptions}
Figure \ref{TxRxAll} shows transmissions and receptions number for all strategies.

\emph{Flooding} and \emph{Pruned Flooding} strategies present the highest transmissions and receptions number. For \emph{Pruned Flooding}, $K$ is the number of nodes to choose randomly. With $K=3$, a node sends three packets which increases the number of packet copies in the network. For \emph{Flooding}, no restrictions on broadcasting a packet, allows to generate many packet copies.
\emph{Plain Flooding} and \emph{CLPB} have a low transmissions and receptions number. However, back to the end to end delay (Figure \ref{EEDAll}) and the percentage of covered nodes (Figure \ref{PercentCrossAll}), \emph{CLPB} performs better. For \emph{Plain Flooding}, this low number is due to lack of communication, while with \emph{CLPB}, this is thanks to scheduling restrictions. 

\begin{figure}[htbp]
 \begin{subfigure}{0.8\columnwidth}
 \centering
 \includegraphics[width=\textwidth,height=4.5cm]{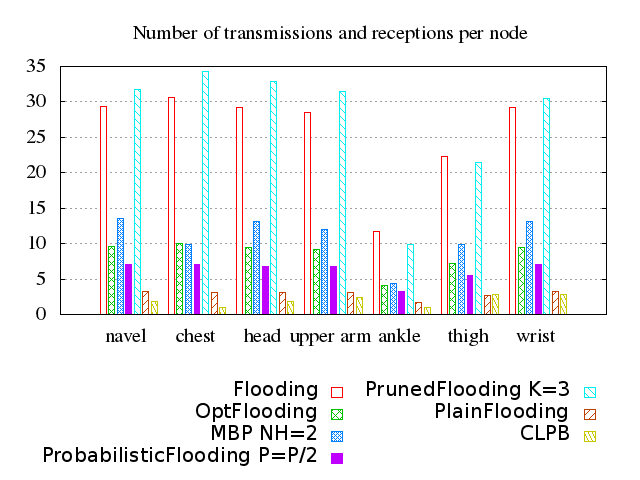}
 \end{subfigure}
 \caption{Transmissions and receptions number per node}
 \label{TxRxNodesAll}
\end{figure}

Figure \ref{TxRxNodesAll} shows transmissions and receptions number per node. 

For all nodes, \emph{CLPB} presents the lowest transmissions and receptions number. Nodes broadcast only once a received packet during their time slot. Thus, the number of transmissions is, at most, equal to $1$ for all nodes. The number of receptions depends on node positions and is at most equal to the number of senders if we suppose that a node is able to intercept all broadcasted packets.
For \emph{Sink} node, there is no receptions. When \emph{Sink} broadcasts the packet in the network, it goes back to sleep and the overall number of transmissions and receptions is equal to $1$.

\begin{figure}[htbp]
 \begin{subfigure}{0.9\columnwidth}
 \centering
 \includegraphics[width=\textwidth,height=4.5cm]{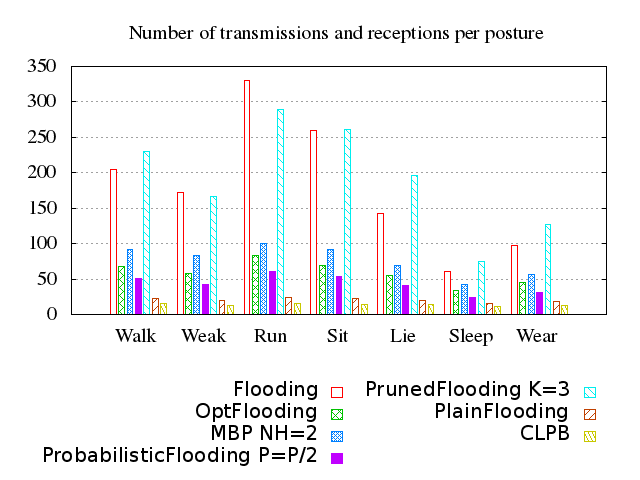}
 \end{subfigure}
 \caption{Total number of transmissions and receptions per posture}
 \label{TxRxPosturesAll}
\end{figure}

Figure \ref{TxRxPosturesAll} shows the average number of transmissions and receptions per posture. 

Results show that our new protocol is the less affected by body posture. Figure \ref{TxRxPosturesAll} shows an almost equal number of transmissions and receptions for all postures. This is explained by a number of transmissions bounded by $1$ at most and a number of receptions bounded by \emph{Number of time slots or in other words number of senders}. 

For the other strategies, the number of transmissions and receptions varies depending on the wearer posture. We can observe the lowest number is with SLEEP posture and the highest number is with RUN posture. We observe also that SIT posture present a high number of transmissions and receptions. This posture is a Up-Down motion variation and is characterized with a dense network where nodes are interconnected. 

\emph{
\emph{CLPB} offers the best of two worlds: high percentage of covered nodes and an end-to-end delay similar to the best known flat strategy (i.e. \emph{Flooding}).
Simulations results also showed $0$ interferences and dropped frames with interference. 
Collisions between packets are reduced to $0$ since at each slot only one node is allowed to transmit. 
However, there are a number of dropped frames without interferences which is due to our mobility model characteristics.}

%%%%%%%%%%%%%%%%%%%%%%%%%%%%%%%%%%%%%%%%%%%%%%%%%%%%%%%%%%%%%%%%%%%%%%%%%%%%%%%%%%%%%%%%%%%%%%%

\section{Conclusion and future works} 
\label{Conclusion}
This paper is, to the best of our knowledge, the first that proposes a MAC-network cross-layer broadcast in WBAN. 
Our work was motivated by results obtained after an extensive set of simulations where we 
stressed the existing network layer broadcast strategies \cite{BCPP15} against realistic human body mobility and various transmission rates. 
With no exception, the existing flat broadcast strategies register a dramatic drop of performances in terms of percentage of covered nodes, end-to-end delay and energy consumption when the transmission rates are superior to $11$Kb/s. 
We therefore, propose a new MAC-network cross-layer broadcast 
that exploits the communication graph defined by the body posture in order to optimize the medium access and nodes synchronization.
Our new protocol outperforms existing flat broadcast strategies in terms of percentage of covered nodes, energy consumption and native correct reception of causally-ordered packets (i.e. packets are received in the order of their sending). Furthermore, our protocol maintains its good performances up to $190$Kb/s transmission rates. 

Our work opens several research directions. In the following we discuss two of them. First, we plan to investigate the slot synchronization in WBAN.The cross-layer protocols designed so far for WBAN assume a strong slot synchronization. Efficiently synchronizing slots in WBAN with realistic human body postures and mobility is an open issue. Second, we intend to extend our study to cross-layer converge-cast protocols. Although, there are several proposals in the WBAN literature, none of them has been stressed with realistic human body mobility and peaks of transmission rates. 

\bibliographystyle{ACM-Reference-Format}
\bibliography{biblioJournal}

\end{document}